\documentclass[aps, floatfix, nofootinbib, amsmath, amssymb, eqsecnum, twocolumn, fleqn]{revtex4}
\usepackage[french]{babel}
\usepackage{color}
\usepackage{graphicx}
\usepackage{float}
\usepackage{pgfplots}
\usetikzlibrary{external}
\usepackage{ifthen}
\usepackage{bm}
\usepackage{upgreek}
\usepackage{hyperref}
\usepackage{soul}
\usepackage[LGR,T1]{fontenc}
\DeclareSymbolFont{upgreek}{LGR}{cmr}{m}{n}
\SetSymbolFont{upgreek}{bold}{LGR}{cmr}{bx}{n}
\usepackage{dutchcal}
\setcitestyle{authoryear,round, sort}

\newcommand{\mf}{\mathcal{f}}
\DeclareMathSymbol{\uptheta}{\mathord}{upgreek}{`j}
\newcommand{\btheta}{{\bm{\uptheta}}}
\renewcommand{\theta}{\vartheta}
\newcommand{\bmu}{{\bm{\mu}}}
\newcommand{\bd}{{\bf d}}
\newcommand{\bC}{{\bf C}}
\newcommand{\bF}{{\bf F}}
\newcommand{\bw}{\bm{w}}
\newcommand{\bb}{\bm{b}}

\newcommand{\bn}{{\bf n}}
\newcommand{\bh}{{\bf h}}
\newcommand{\br}{{\bf r}}
\newcommand{\bs}{{\bf s}}
\newcommand{\ba}{\bm{a}}
\newcommand{\bx}{{\bf x}}

\renewcommand{\(}{\left(}
\renewcommand{\)}{\right)}
\newcommand{\llanglel}{\left.\left\langle}
\newcommand{\rranglel}{\right\rangle\right|}

\definecolor{blue}{HTML}{1F77B4}
\definecolor{orange}{HTML}{FF7F0E}
\definecolor{green}{HTML}{2CA02C}

\newcommand{\hMpc}{h\,\text{Mpc}^{-1}}

\begin{document}
	\author{Tom Charnock}
		\email{charnock@iap.fr}
		\affiliation{Sorbonne Universit\'e, CNRS, UMR 7095, Institut d'Astrophysique de Paris, 98 bis boulevard Arago, 75014 Paris, France}
	\author{Guilhem Lavaux}
		\email{lavaux@iap.fr}
		\affiliation{Sorbonne Universit\'e, CNRS, UMR 7095, Institut d'Astrophysique de Paris, 98 bis boulevard Arago, 75014 Paris, France}
		\affiliation{Sorbonne Universit\'es, Institut Lagrange de Paris, 98 bis boulevard Arago, 75014 Paris, France}
    \author{Benjamin D. Wandelt}
		\email{wandelt@iap.fr}
		\affiliation{Center for Computational Astrophysics, Flatiron Institute, 162 5th Avenue, 10010, New York, NY, USA}
		\affiliation{Sorbonne Universit\'e, CNRS, UMR 7095, Institut d'Astrophysique de Paris, 98 bis boulevard Arago, 75014 Paris, France}
        \affiliation{Sorbonne Universit\'es, Institut Lagrange de Paris, 98 bis boulevard Arago, 75014 Paris, France}
        \affiliation{Department of Astrophysical Sciences, 4 Ivy Lane, Princeton University, Princeton, NJ 08544, USA}
        %\affiliation{Department of Physics and Astronomy, University of Illinois at Urbana-Champaign, 1002 W Green St, Urbana, IL 61801, USA}
			
    \title{Automatic physical inference with information maximising neural networks}
	\begin{abstract}
    Compressing large data sets to a manageable  number of  summaries that are informative about the underlying parameters vastly simplifies both frequentist and Bayesian inference. When only simulations are available, these summaries are typically chosen heuristically, so they may inadvertently miss important information. We introduce a simulation-based machine learning technique that trains artificial neural networks to find non-linear functionals of data that maximise Fisher information: information maximising neural networks (IMNNs). In test cases where the posterior can be derived exactly, likelihood-free inference based on  automatically derived IMNN summaries produces nearly exact posteriors, showing that these summaries are good approximations to sufficient statistics. In a series of numerical examples of increasing complexity and astrophysical relevance we show that IMNNs are robustly capable of automatically finding optimal, non-linear summaries of the data even in cases where linear compression fails: inferring the variance of Gaussian signal in the presence of noise; inferring cosmological parameters from mock simulations of the Lyman-$\alpha$ forest in quasar spectra; and inferring frequency-domain parameters from LISA-like detections of gravitational waveforms. In this final case, the IMNN summary outperforms linear data compression by avoiding the introduction of spurious likelihood maxima. We anticipate that the automatic physical inference method described in this paper will be essential to obtain both accurate and precise cosmological parameter estimates from complex and large astronomical data sets, including those from LSST and Euclid.
\end{abstract}
	\maketitle

	Current data analysis techniques in astronomy and cosmology often involve reducing large data sets into a collection of sufficient statistics~\citep{Bond:1998, Tegmark:1996bz, Heavens:1999am}.
    There are several methods for condensing raw data to a set of summaries.
    Amongst others, these methods could be: principal component analysis (PCA)~\citep{Heck:1987, Francis:1992, Connolly:1994ph, Madgwick:2002, Lahav2009}; statistics including the mean, covariance, and higher point functions~\citep{Belmon:2002,babu2012statistical_wegman} or; calculating the autocorrelation or power spectrum~\citep{babu2012statistical_wegman, babu2012statistical_segal}.
	Unfortunately, summaries calculated using the above methods can still be infeasibly large for data-space comparison.
	For example, analysis of weak lensing data from the Euclid and the Large Synoptic Survey Telescope (LSST) photometric surveys will have around $10^4$ summary statistics~\citep{Heavens:2017efz}.
     Reducing the number of summaries further results in enormous losses in the information available in the raw data~\citep{Heavens:2017efz}.
     
     \medskip
    Another popular way of summarising data is using the Massively Optimised Parameter Estimation and Data (MOPED) compression algorithm~\citep{Heavens:1999am}.
    Summaries from MOPED are linear combinations of data that compress the number of data points down to the number of parameters of a model describing the data.
    MOPED is completely lossless when noise in the data is independent of the parameters and when the likelihood is, at least to first order, Gaussian~\citep{Heavens:1999am}.
    The MOPED algorithm has been used on many problems in astronomy and cosmology such as studying the star formation histories of galaxies~\citep{Reichardt:2001ku, Heavens:2004,Panter:2006mg}, analysing the cosmic microwave background~\citep{Gupta:2002, Zablocki:2016}, and identifying transients~\citep{Protopapas:2005ws} to name but a few.
	Unfortunately, using linear combinations of the data for compression may not be optimal for maximising the possible information available, even when the likelihood is known~\citep{Alsing:2017var}.
    
    \medskip
    For many astronomical and cosmological problems, it can become impossibly difficult to write a likelihood function which describes, not only physics, but also includes any selection bias and instrumental effects.
    Recently, methods have become available to perform inference when a likelihood is not available via approximate Bayesian computation (ABC).
    ABC is a technique which allows samples to be drawn from an approximate posterior distribution.
    Forward simulations are first created using parameter values drawn from a prior and samples are accepted or rejected by comparing the \emph{distance} of the simulation to the real data.
    To efficiently approach the true posterior distribution, it is convenient to couple ABC with a sampling procedure such as population Monte Carlo (PMC).
    ABC using PMC (PMC-ABC) is a method to obtain approximate parameter distributions by iterating through weighted samples from the prior~\citep{Pritchard:1999, Tavere:1997} and can massively reduce the number of samples which need to be drawn during ABC. 
    
    \medskip
    Likelihood-free inference has been used for a variety of astronomical problems which include deducing quasar luminosity functions~\citep{Schafer:2012}, understanding early time galaxy merger rate evolution~\citep{Cameron:2012}, constraining cosmological parameters with supernova observations~\citep{Weyant:2013}, interpreting galaxy formation~\citep{Robin:2014}, searching for the connection between galaxies and halos~\citep{Hahn:2016zwc}, measuring cosmological redshift distributions~\citep{Kacprzak:2017nxl}, inferring photometric evolution of galaxies~\citep{Carassou:2017}, and calculating the ionising background using the Lyman-$\alpha$ and Lyman-$\beta$ forest transmission~\citep{Davies:2017eir}.
    Each of the above examples are used in conjunction with publicly available (PMC-)ABC codes~\citep{Ishida:2015wla, Akeret:2015, Jennings:2016ibb}.
    
    \medskip 
	A two-step compression algorithm was defined in~\citep{Alsing:2018eau} that is capable of optimally summarising data whilst preserving information when the likelihood is not known.
    The first step involves extracting informative statistics from raw data (or simulations of the data) heuristically, i.e. perhaps using the power spectrum or using PCA.
    The summaries of the simulations contain information about physics, selection bias and the instrument.
    A second step then assumes an asymptotic likelihood to perform compression from the summaries gathered in the first step down to the number of parameters in the model as in MOPED or~\citep{Alsing:2017var}.
    The choice of likelihood in the second step does not bias the inference of model parameters during ABC, although the compression will be closer to optimal by choosing a better likelihood function.
    
    \medskip
    However, what if there is information in the data that we did not think to summarise in a first-step summary?
    In this paper we introduce the concept of \emph{information maximising neural networks} (IMNNs).
    Through the use of machine learning, we can circumvent the two step compression used in~\citep{Alsing:2018eau} and find the most informative non-linear data summaries by training a neural network using the Fisher information matrix as a reward function.
    In fact, if we already know some informative summaries, such as those calculated in the first step of~\citep{Alsing:2018eau}, we can use the IMNN to calculate  summaries of the data which optimally increase the information further and then including the IMNN summaries amongst the first-step summaries.

	\medskip
	Once the network is trained, ABC proceeds as before.
    Model parameters can be drawn from a prior, used to generate simulations and once they are fed through the network, the IMNN summaries of the simulation can be compared to the  summaries of the real data.
    Samples can then be accepted or rejected given the distance of the network summary of the simulation to the network summary of the real data to build the approximate posterior distribution of model parameters.
	The IMNN provides a framework to perform automatic physical inference simply by producing simulations.

\medskip
	In section~\ref{sec:MOPED} we describe how to calculate the Fisher information matrix and how linear summaries of the data can conserve Fisher information using the MOPED algorithm.
    In section~\ref{sec:fisher} we lay out the procedure for creating non-linear summaries of the data.
	An overview of how artificial neural networks work is presented in section~\ref{sec:ANN} and we continue in section~\ref{sec:FAAN} by showing how maximising the determinant of the Fisher information matrix allows a network to be trained to provide the optimal non-linear set of summaries.
	Next, in section~\ref{sec:ABC}, we trace the steps to obtain parameter constraints from PMC-ABC using the network trained as prescribed in section~\ref{sec:FAAN}.
	Finally, in section~\ref{sec:test}, we give some test examples.
	The first test model provides an example where a single linear summary of the data would provide nearly no information about a parameter, but the non-linear summary provided by a trained artificial neural network can extract the maximum information the data contains.
	The second example is more astronomically motivated, using the absorption of flux from quasars by neutral hydrogen to constrain the amplitude of scalar perturbations.
    Finally we use the network to summarise and constrain the central oscillation frequency of a gravitational wave burst from Laser Interferometer Space Antenna (LISA).
    This problem was used in~\citep{Graff:2010dt} to show that MOPED compression introduces spurious maxima in the posterior distribution; we show that the non-linear IMNN  data compression introduced in this paper can avoid this peculiarity.
	  
	\section{Fisher information and linear compression\label{sec:MOPED}}
	
	A likelihood function $\mathcal{L}\(\bd|\btheta\)$ of some data, $\bd$, with $n_\bd$ data points, is informative about a model with a set of $n_\btheta$ parameters, $\btheta$.
    The more sharply peaked $\mathcal{L}\(\bd|\btheta\)$ is at a particular value of $\btheta$, the better $\btheta$ is known.
	The Fisher information describes how much information $\bd$ contains about the linear parameters, $\btheta$, and can be calculated by finding the second moment of the score of the likelihood~\citep{Fisher:1925, Kenney:1951, Kendall:1969}, i.e. the variance of the partial derivative of the natural logarithm of the likelihood with respect to the parameters at a fiducial parameter value, $\btheta^{\rm fid}$,
    \begin{align}
		\bF_{\alpha\beta}\(\btheta\) &=\int \mathrm{d}\bd\,\mathcal{L}\(\bd|\btheta\)\left.\frac{\partial\ln\mathcal{L}(\bd|\btheta)}{\partial\theta_\alpha}\frac{\partial\ln\mathcal{L}(\bd|\btheta)}{\partial\theta_\beta}\right|_{\btheta=\btheta^{\rm fid}}\nonumber\\
		&=\llanglel\frac{\partial\ln\mathcal{L}\(\bd|\btheta\)}{\partial\theta_\alpha}\frac{\partial\ln\mathcal{L}\(\bd|\btheta\)}{\partial\theta_\beta}\rranglel_{\btheta\,=\,\btheta^{\rm fid}}.\label{e:F1}
	\end{align}	
	Equation~\eqref{e:F1} can be rewritten as
	\begin{align}
		\bF_{\alpha\beta}\(\btheta\) &= -\llanglel\frac{\partial^2\ln\mathcal{L}\(\bd|\btheta\)}{\partial\theta_\alpha\partial\theta_\beta}\rranglel_{\btheta\,=\,\btheta^{\rm fid}}\label{e:fisher}
	\end{align}
	when the likelihood is twice continuously differentiable~\citep{Kenney:1951, Kendall:1969, Lehmann:2003}.
	A large Fisher information for a given set of data indicates that the data is informative about the parameters and therefore the parameters can be measured more effectively~\citep{Lehmann:2003}.
	In particular, the minimum variance of an estimator of a parameter, $\btheta$, is given by the Cram\'er-Rao bound~\citep{Cramer:1946, Rao:1945}, which states that
	\begin{align}
		\langle(\theta_\alpha-\langle\theta_\alpha\rangle)(\theta_\beta-\langle\theta_\beta\rangle)\rangle &\geq \(\bF^{-1}\)_{\alpha\beta},
	\end{align}
	such that finding the maximum Fisher information, provides the minimum variance for estimators of $\btheta$.
    Note that the Cram\'er-Rao inequality only holds under certain conditions, i.e. that the score function is defined for all $\bd$ in the support of the likelihood and that differentiation and taking the expectation commute.
    The Cram\'er-Rao bound limits the second moment of any estimator, but does not limit the shape of the confidence regions~\citep{Sellentin:2014zta}.
    In the case that the likelihood of the data in a particular model is Gaussian, the logarithm of the likelihood can be written as
	\begin{align}
		-2\ln\mathcal{L}(\bd|\btheta) &= (\bd-\bmu(\btheta))^T\bC^{-1}(\bd-\bmu(\btheta))+\ln\left|2\pi\bC\right|,\label{e:likelihood1}
	\end{align}
	where $\bd$ is the data and $\mu(\btheta)$ is the mean of the model given parameters $\btheta$, which we will denote $\bmu$ for convenience. 
    $\bC$ is the covariance of the data and is assumed to be independent of the parameters.
	Using the MOPED algorithm~\citep{Heavens:1999am}, $\bd$ can be compressed from the number of points in the data, $n_\bd$, to the number of parameters of the model, $n_\btheta$, simply by seeking the linear combination of data which optimises the linearised parameters.
	The MOPED compression is lossless in the sense that the Fisher information is conserved under the transformation
    \begin{align}
    	x_\alpha & = \br_\alpha^T\bd\label{e:MOPED}
    \end{align}
    where $\alpha$ labels the parameter and $\br_\alpha$ is calculated by maximising the Fisher information ensuring that $\br_\alpha$ is orthogonal to $\br_\beta$ (where $\alpha\ne\beta$).
    The form of  $\br_\alpha$ is
    \begin{align}
    	\br_1 & = \frac{\bC^{-1}\bmu,_1}{\sqrt{\bmu,_1^T\bC^{-1}\bmu,_1}},\label{e:r1}
    \end{align}
    for the first parameter, $\theta_1$, and where $\partial/\partial\theta_\alpha\equiv\,\,\,,_{\alpha}$.
    For each parameter afterwards,
    \begin{align}
    	\br_{\alpha} = & \frac{\bC^{-1}\bmu,_\alpha-\sum_{i=1}^{\alpha-1}\(\bmu,_\alpha^T\br_i\)\br_i}{\sqrt{\bmu,_\alpha^T\bC^{-1}\bmu,_\alpha-\sum_{i=1}^{\alpha-1}\(\bmu,_\alpha^T\br_i\)^2}}.
    \end{align}
    After creating the linear summaries, $\bx = \{x_\alpha|\,\alpha\in[1,\,n_\btheta]\}$, $\bx$ is as informative about $\btheta$ as $\bd$ is with regards to the Fisher information, for the likelihood in equation~\eqref{e:likelihood1}.
	The Fisher information takes the form
    \begin{align}
		\bF_{\alpha\beta} & = {\rm Tr}\left[\bmu,_\alpha^T\bC^{-1}\bmu,_{\beta}\right],\label{e:gaussianfisher}
	\end{align}
	The lossless compression of the data, $\bd\to\bx$, is only possible when the likelihood is exactly of the form in equation~\eqref{e:likelihood1}.
	Nearly lossless compression is still possible if the peak of the likelihood is approximately Gaussian. Often, this will be a good approximation in the asymptotic limit, i.e., when  the data are informative about the parameters.

	\section{Non-linear Fisher information maximising summaries\label{sec:fisher}}

	We are influenced by the MOPED algorithm to find some transformation which maps the data to compressed summaries, $\mf:\bd\to\bx$, whilst conserving Fisher information, but without the limitation that the method is only valid as a Gaussian approximation.
    $\mf$ is a function that modifies the original likelihood describing the data, which need not be known \emph{a priori}, into the form
    \begin{align}
        -2\ln\mathcal{L}\(\bx|\btheta\) &=\(\bx - \bmu_\mf\(\btheta\)\)^T\bC_\mf^{-1}\(\bx - \bmu_\mf\(\btheta\)\)\label{e:fgaussian}
    \end{align}
    where 
	\begin{align}
        \bmu_\mf(\btheta) & = \frac{1}{n_\bs}\sum_{i=1}^{n_\bs}\bx^\bs_i\label{e:fmu},
    \end{align}
	is the mean value of $n_\bs$ summaries, $\{\bx^\bs_i|\,i\in[1,\,n_\bs]\}$, where each summary is obtained from a simulation $\bd^\bs_i = \bd^\bs(\btheta, i)$ using $\mf:\bd^\bs_i\to\bx^\bs_i$.
    We will denote $\mu_\mf(\btheta)\equiv\bmu_\mf$ for convenience.
    Each $i$ denotes a different random initialisation of a simulation.
    Similarly $\bC^{-1}_\mf$ is the inverse of the covariance matrix which is again obtained from simulations of the data
    \begin{align}
        \(\bC_{\mf}\)_{\alpha\beta} = &
          \frac{1}{n_\bs-1}
          \sum_{i=1}^{n_\bs}\(\bx^\bs_{i}-\bmu_{\mf}\)_\alpha\(\bx^\bs_{i}-\bmu_{\mf}\)_\beta.\label{e:fcov}
    \end{align}
    Using equation~\eqref{e:fisher} a modified Fisher information matrix can be calculated from the likelihood in equation~\eqref{e:fgaussian}
    \begin{align}
        \bF_{\alpha\beta}&= {\rm Tr}\left[\bmu_\mf,_\alpha^T\bC_\mf^{-1}\bmu_\mf,_\beta\right].\label{e:wbfisher}
    \end{align}
	Here, the values of $\bmu_\mf,_\alpha$ and $\bC^{-1}_\mf$ are calculated using fixed, fiducial parameter values, $\btheta^{\rm fid}$, such that the simulations are $\bd_i^{\bs~{\rm fid}}=\bd^\bs(\btheta^{\rm fid}, i)$.
     Although $\mf:\bd\to\bx$ is not specified, a subclass of $\mf$ is accessible via a neural network, described in detail in section~\ref{sec:ANN}.
    We will show how this function can be found by training a neural network in section~\ref{sec:FAAN}.

    \section{Artificial neural networks\label{sec:ANN}}
    
	Artificial neural networks are arbitrary maps from some inputs to outputs.
	Consider some data vector $\bd=\left\{d_i \big|\,i \in[1,\,n_\bd]\right\}$ with $n_\bd$ data points.
	Each data point is regarded as an input to a network.
	For a deep neural network, a series of \emph{hidden layers} are able to learn levels of abstraction from the input~\citep{Bengio:2009, Cybenko:1989, Deng:2014, Goodfellow:2016, Nielsen:2015}.
	Each layer, $l$, of the network contains a set of \emph{neurons} which takes some number of inputs and provides one output per neuron~\citep{McCulloch:1943, Pitts:1947}.
	The output a neuron is \emph{activated} by a non-linear activation function 
    \begin{align}
    	a_i^l &= \phi\(v_i^l\)\label{e:activa}
    \end{align}
    where
	\begin{align}
		v^l_j &= \sum_i w^l_{ji}a_i^{l-1}+b^l_j,\label{e:wbi}
	\end{align}%
	is a weighted, biased input at each layer with weights $\bw^l \equiv w_{ji}^l$ and biases $\bb^l \equiv b^l_j$~\citep{McCulloch:1943}. 
    $i$ describes an element of the output vector of a collections of neurons in the $(l-1)^{\rm th}$ layer and $j$ indexes the neuron in layer $l$.
    With these notations, the input to the network can be considered to be the output of a zeroth layer of a network, $d_i \equiv a_i^0$.
	Stacking several neurons into a hidden layer and stacking several hidden layers, taking the outputs from the previous layer as the inputs to each node in the next layer, allows for greater levels of abstraction from the input data~\citep{Deng:2014}.
    These networks are often referred to as \emph{deep} networks.
    Note that the addition of too many layers can lead to expensive computations and overfitting by the network so that it becomes difficult to train.
	The network output at the final layer can be described by $\ba^L = \{a_i^L|\,i \in [1,\,n_{\rm outputs}]\}$ where $n_{\rm outputs}$ is the number of outputs in the final layer, labelled $L$, and $a^L_i=\phi(v_i^L)$.
    
    \medskip
    As mentioned at the end of section~\ref{sec:fisher}, a neural network can be used as a representation of $\mf:\bd\to\bx$, which compresses data to summary statistics.
    Formally, this subclass of functions is described, for some input ${\bf z}$, by
    \begin{align}
    	\mf^l: {\bf z} \to \ba^l &= \phi\left(\sum_i w^l_{ji} \left[\mf^{l-1}({\bf z})\right]_i + b^l_j\right),
    \end{align}
    for $l > 0$ and
    \begin{align}
    	\mf^0: {\bf z} &\to \ba^0 = {\bf m},
    \end{align}
    where the compressed summary is given at $l=L$ of the recursion and the input to the function at $l=0$ is taken to be the identity.
    
    \subsubsection{Activation functions}
    
    The activation function, $\phi(v_i^l)$, in equation~\eqref{e:activa} describes whether the artificial neuron \emph{fires} or not, i.e. whether the inputs are informative or useful for describing the output~\citep{Cybenko:1989, He:2015, Krizhevsky:2012, Nielsen:2015}.
    It is the activation function that provides the non-linearity necessary for the the network to learn the complex map from inputs to outputs by combining the relevant combinations of inputs at each layer in a non-trivial way.
    As long as there are enough hidden layers, the form of the activation function is relatively unimportant since the weights and biases will be trained to combine the outputs of each hidden layer in such a way as to provide the correct map.
    There are many options for the choice of activation function, including $\tanh$ and {\ttfamily sigmoid} functions.
	Currently popular activation functions are the rectified linear unit ({\ttfamily ReLU})~\citep{He:2015}.
	We show here, as an example, an adaptation called leaky {\ttfamily ReLU}
    \begin{align}
        \phi\(x\) &= \left\{\begin{array}{ll}\alpha x & \b x\le 0\\\phantom{\alpha}x & x> 0\end{array}\right.,
    \end{align}%
    where $\alpha = 0$ for {\ttfamily ReLU} and $\alpha$ is small and positive for leaky {\ttfamily ReLU}~\citep{Maas:2013}.
    Although the {\ttfamily ReLU} family of activation functions are linear, stacking several layers of neurons provides a function which approximates a non-linear function, and is extremely quick to calculate.
    It will become apparent that the derivative of the activated output with respect to the weighted, biased inputs are essential for training neural networks.
    The derivative of the {\ttfamily ReLU} family of activation functions can also be efficiently calculated as
    \begin{align}
        \frac{\partial\phi\(x\)}{\partial x} &= \left\{\begin{array}{ll}\alpha & x\le 0\\1 & x> 0\end{array}\right..\label{e:dphidv}
    \end{align}
    Although we have shown {\ttfamily ReLU} as an example, we explore various activation functions across the population of networks that we train.
    
    \subsubsection{Back propagation}
	
	A scalar loss function, $\Lambda(\ba^L)$, is calculated from the outputs of the network $\ba^L$.
    In supervised deep learning, the loss function describes how far the outputs are from a set of labels for the training data~\citep{Rumelhart:1986}.
    An iterative procedure, called back propagation, uses the chain rule to find how much the weights and biases need to change to minimise the loss function~\citep{Rumelhart:1986}.
    Using gradient descent~\citep{Kiwiel2001} it can be seen that the weights and biases must be updated using
    \begin{align}
    	w_{ji}^l&\to w_{ji}^l-\eta\frac{\partial \Lambda}{\partial w_{ji}^l}\label{eq:wupdate}
    \end{align}%
    and
    \begin{align}
    	b_i^l&\to b_i^l-\eta\frac{\partial \Lambda}{\partial b_i^l},\label{eq:bupdate}
    \end{align}%
    where $\eta$ is a tunable learning rate which dictates the size of the steps that the weights and biases are able to take on each update~\citep{Nielsen:2015}.
    It is very efficient to calculate the derivatives in equations~\eqref{eq:wupdate} and~\eqref{eq:bupdate} at the last layer using
    \begin{align}
    	\frac{\partial \Lambda}{\partial v_i^L} & = \frac{\partial \Lambda}{\partial a_i^L}\frac{\partial a_i^L}{\partial v_i^L}.\label{eq:dldv}
    \end{align}
    From any layer, the rate of change of the loss function with respect to the weighted, biased inputs at the previous layer can be found using
   	\begin{align}
   		\frac{\partial \Lambda}{\partial v_i^l} & = \sum_{j}w_{ji}^{l+1}\frac{\partial \Lambda}{\partial v_j^{l+1}}\frac{\partial a_i^l}{\partial v_i^l}.
	\end{align} 
	The changes in the loss function under changes in the weights or the biases are then calculated using
    \begin{align}
    	\frac{\partial \Lambda}{\partial w_{ji}^l} & = \frac{\partial \Lambda}{\partial v_j^l}\frac{\partial v_j^l}{\partial w_{ji}^l}\nonumber\\
    	& = \frac{\partial \Lambda}{\partial v_j^l}a_i^{l-1}\label{eq:dldw}
    \end{align}
    and 
    \begin{align}
    	\frac{\partial \Lambda}{\partial b_i^l} & = \frac{\partial \Lambda}{\partial v_i^l}\frac{\partial v_i^l}{\partial b_i^l}\nonumber\\
    	& = \frac{\partial \Lambda}{\partial v_i^l}.\label{eq:dldb}
   	\end{align}%
   	Each of the $a_i^l$ and the derivatives with respect to the weighted biased inputs (using equation~\eqref{e:dphidv}) are calculated on the forward pass of the network inputs.
   	Back propagation allows the change of the loss function with respect to all of the weights or biases to be calculated in just one pass forward and one pass backwards~\citep{Nielsen:2015}.
    By calculating the change in the loss function with respect to the network outputs and applying equation~\eqref{eq:dldv} successively, the weight and bias updates at every layer can be calculated easily.
    
    \medskip
    The back propagation procedure is repeated many times using different sets of training inputs~\citep{Nielsen:2015}.
    Once all of the training inputs are used, one epoch of training is complete.
    After one epoch of training, the order of the training inputs can be jumbled and the training procedure repeated many times until the loss function is minimised~\citep{Nielsen:2015}.
        
    \subsubsection{Overfitting}
	It is possible that the network weights become tuned to features in the training data which are not present in the real data.
    To prevent this \emph{overfitting}, we implement \textit{dropout} ~\citep{Srivastava:2014}.
    Dropout is a technique where a random fraction of the neurons are set to zero on each batch of training and after back propagation only the weights and biases of the active neurons are updated.
    Performing dropout during training equates to training many sub-networks, where all the neurons share weights and biases.
    Each of the sub-networks can learn specific features in the data, but the consensus network does not learn features too strongly.

	\subsubsection{Training and test data sets}
	When training a network, it is essential to test how well the network is learning by using a \emph{test} set which contains data which is not present in the training set.
    However, it is extremely important to note that even the accuracy of prediction on the \emph{test} set should not be considered to be a measure of the predictive ability of the network.
    It is considered normal to tune a network to achieve the minimal loss of the test set without showing signs of overfitting.
    A \emph{third}, completely unseen, data set should then be used to quote network accuracies.
    In doing so, the irreproducable accuracy scores often quoted in the literature, arising from only considering a network that is highly tuned on the test set, are avoided.
    In this paper we train and test networks with a training set and a test set and use the comparison between the posterior distribution obtained using the network output and the analytically calculated distribution as our confirmation that the network is accurate.

    	\section{Finding non-linear summaries\label{sec:FAAN}}

    Inspired by supervised artificial neural networks we are able to create a network capable of maximising the Fisher information to create non-linear summaries of data.
    The output of the network, $\bx\equiv\ba^L$ is a compressed summary of some data, $\bd$.
    Since the data is a function of some parameters, $\btheta$, given some model, the summary can be described as a function of these parameters, as well as the weights and biases at each layer, $l$, of a network, $\bx\to \bx\(\btheta, \bw^l, \bb^l\)$.
    The mean, $\bmu_\mf$, covariance, $\bC_\mf$ and Fisher information matrix, $\bF_{\alpha\beta}$, from equations~\eqref{e:fmu},~\eqref{e:fcov} and~\eqref{e:wbfisher}, each become functions of the weights and biases as well.
    Summaries of simulations, $\bx^\bs_i$, are obtained by passing simulations, $\bd^\bs_i$, through the network $\mf:\bd^\bs_i\to\bx^\bs_i$.

\medskip
    To compute the Fisher information matrix in equation~\eqref{e:wbfisher}, the derivative of the network needs to be calculated with respect to the parameters at fiducial values.
    It is, in principle, simple to find the derivative of the network with respect to the parameters due to partial derivatives commuting with sums
    \begin{align}
        \bmu_\mf,_{\alpha} &= \frac{\partial}{\partial\theta_\alpha}\frac{1}{n_\bs}\sum_{i=1}^{n_\bs}\bx^{\bs\,{\rm fid}}_i\nonumber\\
        &=\frac{1}{n_\bs}\sum_{i=1}^{n_\bs}\left(\frac{\partial\bx}{\partial\theta_\alpha}\right)_i^{\bs\,{\rm fid}}.\label{e:dmudt}
    \end{align}
    Unfortunately, since the parameters only appear in the simulations, numerical differentiation needs to be performed.
    The numerical differentiation is achieved by producing three copies of the simulation, $\bd^{\bs\,{\rm fid}}_i=\bd^\bs\(\btheta^{\rm fid},i\)$, $\bd^{\bs\,{\rm fid}-}_i=\bd^\bs\(\btheta^{\rm fid}-\Delta\btheta^-,i\)$, and $\bd^{\bs\,{\rm fid}+}_i=\bd^\bs\(\btheta^{\rm fid}+\Delta\btheta^{+},i\)$ where $\Delta\btheta^{\pm}$ is some small deviation from the fiducial parameter value.
    The derivative of the network output with respect to the parameters is therefore given by
    \begin{align}
        	\left(\frac{\partial \bx}{\partial\theta_\alpha}\right)^{\bs\,{\rm fid}}_i &\approx \frac{\bx^{\bs\,{\rm fid}+}_i - \bx^{\bs\,{\rm fid}-}_i}{\Delta\theta_\alpha^+ - \Delta\theta_\alpha^-}.\label{e:diff}
    	\end{align}
    Setting the random seed, $i$, to the same value when generating $\bd^{\bs\,{\rm fid}-}_i$ and $\bd^{\bs\,{\rm fid}+}_i$ suppresses the sample variance in estimates of the derivative of the mean.
    Although the network output can vary a lot between different simulations, the derivative with respect to parameters is much more stable to changes in the parameter value, meaning relatively few extra simulations ($n_{\partial\theta} < n_\bs$) need to be computed to calculate the gradient of the mean.

\medskip
    Another way of calculating the derivative of the mean of the network output is to calculate the adjoint gradient of the simulations, and calculate the derivative of the network with respect to the simulations
	\begin{align}
    	\bmu_{\mf},_\alpha &=\frac{1}{n_\bs}\sum_{i=1}^{n_\bs}\sum_{k=1}^{n_\bd} \frac{\partial x^{\bs\,{\rm fid}}_{ik}}{\partial d_k}\frac{\partial d^{\bs\,{\rm fid}}_{ik}}{\partial\theta_\alpha},
	\end{align}
    where $i$ labels the random initialisation of the simulation and $k$ labels the data point in the simulation.
    In certain situations, calculating the adjoint gradient of the simulations may be more efficient than the method described in equations~\eqref{e:dmudt} and~\eqref{e:diff}.

\begin{figure*}
    \centering
    \includegraphics{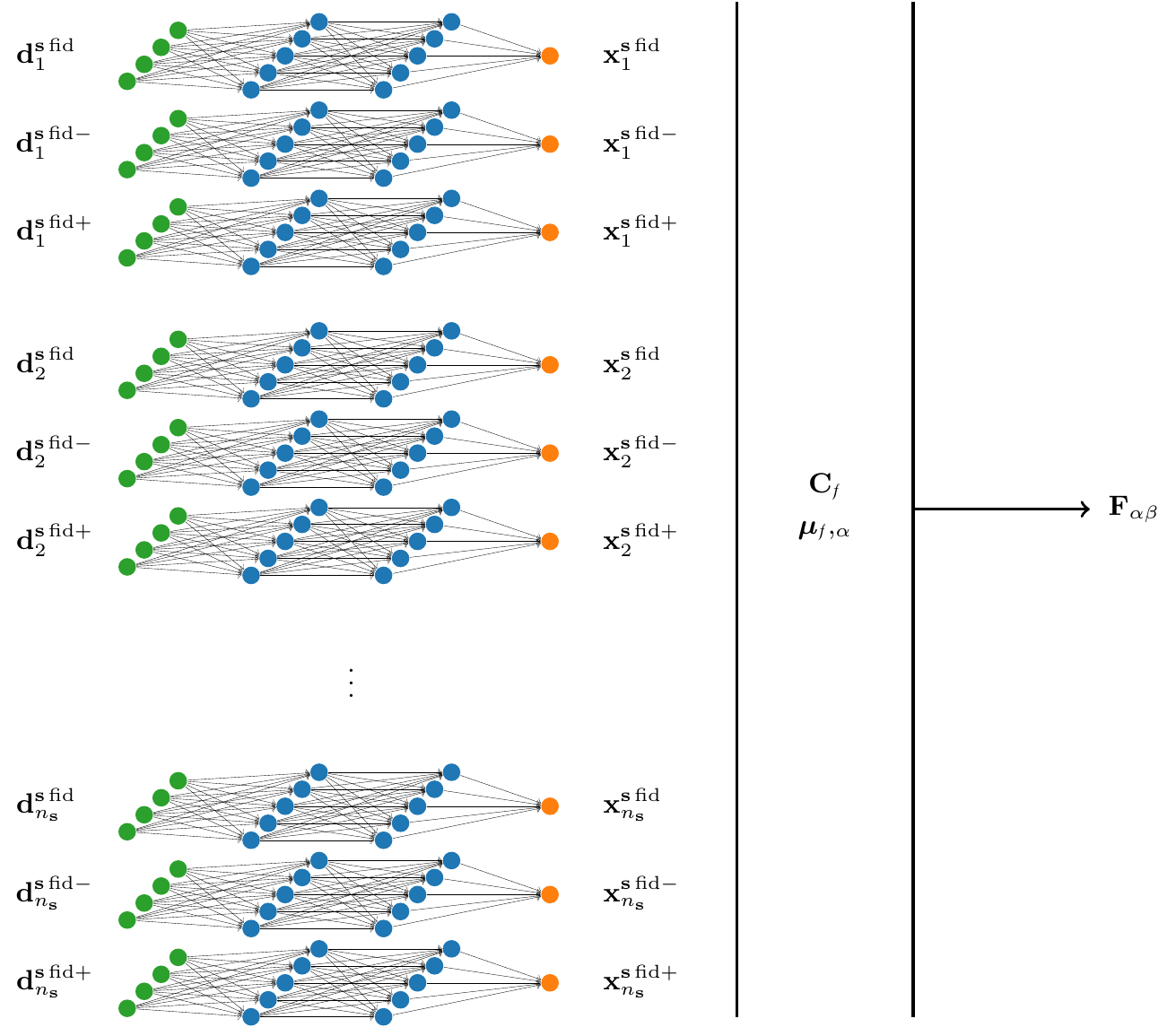}
		\caption{Cartoon of the information maximising neural network architecture.
		During training, each simulation $\bd^{\bs,{\rm fid}}_i$ and each simulation made with a varied fiducial parameter, $\bd^{\bs\,{\rm fid}\pm}_i$, is passed through the same network (all the weights and biases are shared).
		The output of the network for each simulation, $\bx^{\bs\,{\rm fid}}_i$, is used to calculate the covariance, $\bC_\mf$, and each of the network outputs from the varied simulations, $\bx^{\bs\,{\rm fid}\pm}_i$, are used to calculate the derivative of the mean, $\bmu_\mf,_\alpha$.
		The network uses $\partial\Lambda/\partial\ba^L = -|\bF|+|\bC_\mf|$ as the error of a reward function which is maximised through back propagation.
        The reward function is back propagated only through a selection of networks which use the simulations created at the fiducial parameter value,  $\bx^{\bs\,{\rm fid}}_i$.
        The weights and biases are updated using the mean of the back propagated error at each weight and bias, $\partial\Lambda/\partial\bw^l$ and $\partial\Lambda/\partial\bb^l$.
        Once trained, a summary of some data can be obtained using a simple artificial neural network with the weights and biases from the training network.}
		\label{fig:net}
    \end{figure*}

\medskip
    One simple way of obtaining the optimal non-linear summary from some data is to maximise the determinant of the Fisher information matrix calculated from the network, $|\bF|$,
    \begin{align}
    	\Lambda &= -\frac{1}{2}|\bF|^2.
    \end{align}
    The Fisher information matrix terms are produced from the second derivatives of the Kullback-Leibler divergence, i.e. the information gain, and is hence directly related to the Shannon entropy~\citep{Kullback:1968}.
    In particular, the Fisher information matrix is the Shannon information of the Gaussian probability distribution function which optimally approximates the likelihood in~\eqref{e:fgaussian} near its peak.
    For this reason, choosing to maximise the determinant of the Fisher information is equivalent to maximising the Shannon information of this distribution.
    The error is then found by taking the derivative of the loss function with respect to the network output.
    Normally $\bx^\bs_i$ would be considered as the network output when the input is a simulation, but since the quantity of interest in our problem is statistically calculated over a large number of network outputs, we follow the cartoon in figure~\ref{fig:net} and use the determinant of the Fisher information matrix as the true network output.
    First, a large number of simulations at fixed fiducial parameter value and random initialisation (as well as the simulations created to calculate the derivative of the mean) are fed forwards through identical networks.
    All the network outputs from the fixed fiducial parameter simulations are used to calculate the covariance as in equation~\eqref{e:fcov}. 
    Meanwhile, the rest of the network outputs are used to find the derivative of the mean with respect to the parameter as in equations~\eqref{e:dmudt} and~\eqref{e:diff}.
    These are combined to give the Fisher information matrix of equation~\eqref{e:wbfisher}.
    If we consider the \emph{true} network output to be $\ba^L=|\bF|$ rather than $\bx^\bs$ then the error can be defined as
    \begin{align}
    	\frac{\partial \Lambda}{\partial\ba^L} &= -|\bF|.\label{e:error2}
    \end{align}%
    Training then commences over many epochs of weight and bias updates until the Fisher information stops increasing.
    In practice, a problem arises when using equation~\eqref{e:error2}, since the Fisher information is invariant under linear scaling of the summary.
    To control the magnitude of the summaries we can artificially induce a scale by adding the determinant of the covariance matrix, $|\bC_\mf|$, to the error function
    \begin{align}
 		\frac{\partial\Lambda}{\partial\ba^L} &=-|\bF| + |\bC_\mf|.\label{e:realloss}
    \end{align}
    The network is penalised when the determinant of the covariance is large.
    When using equation~\eqref{e:realloss} the network provides the summary which maximises the Fisher information whilst minimising the covariance of the outputs.
    
\medskip
    Although the network is capable of extracting all necessary summaries of the data without any prior knowledge of what the parameters represent, we can imagine the IMNNs would be better suited to extending the heuristic first-step summaries.
    For example, if the power spectrum is a known useful summary of some data, the network can be trained to find any statistic which increases the Fisher information further.
    With the power spectrum and the network summary, a second stage compression as described in~\citep{Alsing:2017var} can be used for efficient parameter inference.
    This way, inexhausted information of the data can be unlocked, even when the form of the data combination that probes it is not known.

    \section{Approximate Bayesian computation\label{sec:ABC}}

	Approximate Bayesian computation (ABC) is a technique of finding an approximate posterior distribution for some model parameters by accepting or rejecting samples dependent on how similar simulations created using the sample parameters are to the real data~\citep{Rubin:1984}.
	It is useful to choose an appropriate sampling procedure to quickly approach the true posterior for the parameters without creating too many simulations.
    Population Monte Carlo (PMC) is an algorithm by which samples can be obtained by iterating through weighted draws from a prior, even when the likelihood is not accessible~\citep{Kitagawa:1996}.
    Although PMC has a variety of uses, such as filtering, we are going to couple it to ABC (PMC-ABC) to effectively approach the true posterior~\citep{Pritchard:1999, Tavere:1997}.
    
    \medskip
    Similar to the method in~\citep{Ishida:2015wla}, our PMC-ABC algorithm starts by drawing $N$ parameter vectors, $\{\btheta_k^{\,t}|\,k \in [1,\, N], t = 0\}$, from the prior, $p(\btheta)$.
    $N$ is the final number of posterior samples wanted, $k$ labels the sample and $t$ describes the number of sampling iterations. 
    In each sampling iteration, samples are drawn from a prior, used to create simulations, and then weighted by the distance of the simulation from the real data.
    The weighted samples are used to obtain a new proposal distribution with which to resample from in the next iteration.
    This allows the PMC-ABC to gradually hone in on the the true probability distribution.
	Simulations are made at each of the $N$ parameter vectors and fed through the trained network to obtain a collection of network summaries $\{\bx^{\bs\,t}_{ik}|\,k\in [1,\, N]\}$ where $i$ labels the simulation.
	Only the value of $\btheta$ is important for ABC and so the random initialisation, $i$, can be ignored once chosen for each simulation.
	We choose to define the distance of each simulated summary from the summary of the real data $\bx$ by
	\begin{align}
		\varrho_k^{t} & = \sqrt{\(\bx^{\bs\,t}_{ik} - \bx\)^T\bF\(\bx^{\bs\,t}_{ik} - \bx\)},\label{eq:distance}
	\end{align}
	where $\bF$ is the Fisher information matrix obtained originally by the network.
    Equation~\eqref{eq:distance} is the optimal distance measure~\citep{Alsing:2017var}, although it is not unique.
	On each iteration, an acceptance condition, $\varepsilon^t$, for the samples is defined by the 75$^{\rm th}$ percentile of $\{\varrho_k^{t}|\,k\in [1,\, N]\}$ such that the 75\% of samples which have the smallest distances from the summary of the real data are kept.
	$\btheta_k^{\,t}$ then corresponds to the remaining 25\% of the samples, which are used to draw parameter vectors for the next iteration, $\btheta_k^{\,t+1}$.
	$\btheta_k^{\,t+1}$ are selected from a Gaussian with mean $\btheta_k^{\,t}$ and covariance, $\bC_t$, from the weighted parameter values.
	The weighting for $\btheta_k^{\,t+1}$ is given by
	\begin{align}
		W_k^{\,t+1} & = \frac{p(\btheta_k^{\,t+1})}{\sum_{j=1}^N W_j^{\,t}\mathcal{N}(\btheta_k^{\,t+1};\btheta_j^{\,t},\bC_t)}
	\end{align}
	with $p(\btheta^{\,t+1}_k)$ as the value of the prior at $\btheta_k^{\,t+1}$,
	\begin{align}
		\mathcal{N}(\btheta_k^{t+1};\btheta_j^{t},\bC_{t}) & = \frac{\exp\left[-\displaystyle\frac{1}{2}\hskip-0.2em\(\btheta^{t+1}_k \hskip-0.2em- \btheta^{t}_j\)^T\hskip-0.35em\bC_t^{-1}\hskip-0.25em\(\btheta^{t+1}_k \hskip-0.2em- \btheta^{t}_j\)\right]}{\sqrt{|2\pi\bC_t|}}
	\end{align}
	and where the initial weighting is equal for all $k$, $W_k^{\,0} = 1/N$.
	$\btheta_k^{\,t+1}$ is drawn repeatedly from the Gaussian with mean $\btheta_k^{\,t}$ and covariance $\bC_t$ until $\varrho_k^{t+1} \le \varepsilon^t$ for each of the rejected $k$ samples.
	Once complete, the first iteration of sampling finishes, allowing $W_k^{\,t+1}$ to be calculated.
    
    \medskip
	Unlike the method in~\citep{Ishida:2015wla} the accepted $\btheta^{\,t}_k$ are instantly promoted to $\btheta_k^{\,t+1}$ rather than being redrawn.
	The accepted $\varrho^{t}_k$ can also be promoted to $\varrho^{t+1}_k$, and the new $\btheta_k^{\,t+1}$ used to find $\bC_{t+1}$.
	The next acceptance condition, $\varepsilon^{t+1}$, is again calculated from the 75$^{\rm th}$ percentile of $\{\varrho^{t+1}_k|\,k\in [1,\, N]\}$ and the selection procedure is repeated.
	Iterations can be performed until the number of draws from $\mathcal{N}(\btheta_k^{\,t},\bC_t)$ in a particular iteration, $t$, is much larger than the number of wanted samples from the posterior, $N$.
	A large number of draws compared to the number of accepted parameter values is a sign that the approximate posterior has stopped changing considerably between iterations.
    
    \section{Testing inference with information maximising neural networks\label{sec:test}}

	In this section we use the information maximising neural network on a range of test models.
	In section~\ref{ss:gaussian} we use the network to summarise a Gaussian signal with unknown variance, as well as Gaussian signal with unknown variance that was contaminated by noise, first of known variance and then of unknown variance.
    We consider the same problem in section~\ref{ss:fid} showing that the network provides nearly optimal, informative summaries in spite of a poorly chosen fiducial parameter value by learning the correct map.
	In section~\ref{ss:lyman} we constrain the amplitude of scalar perturbations using simulations of quasar absorption spectra which can be summarised by a single statistic provided by the network.
	Finally, in section~\ref{ss:graff}, we demonstrate the performance of IMNN compression for the case estimating the central frequency of a LISA gravitational wave chirp. This example  addresses a concern raised in~\citep{Graff:2010dt} where the authors show that a linear summary of data in the time domain can be misleading about a parameter in the frequency domain. We are  show that the non-linear summary avoids this problem and is more informative.
     
	\subsection{Summarising Gaussian signals\label{ss:gaussian}}
	
	A simple toy model can be constructed where linear combinations of the data are unable to provide information about parameters.
    
    \medskip
	Consider an experiment which measures $n_{\bd} = 10$ data points which are drawn from a zero-mean Gaussian where the variance, $\theta = \sigma^2$, is not perfectly known, $\bd = \left\{d_i \curvearrowleft \mathcal{N}\(0, \theta\)\big|\,i \in [1,\,n_{\bd}]\right\}$. 
    The likelihood is written
    \begin{align}
        \mathcal{L}\(\bd|\theta\) &= \prod_{i=1}^{n_\bd}\frac{1}{\sqrt{2\pi\theta}}\exp\left[-\frac{1}{2\theta}d_i^2\right]\nonumber\\
        &=\frac{1}{\(2\pi\theta\)^{n_\bd/2}}\exp\left[-\frac{1}{2\theta}\sum_{i=1}^{n_\bd}d_i^2\right],\label{e:likelihood}
    \end{align}%
    such that
    \begin{align}
        -2\ln\mathcal{L}\(\bd|\theta\) &= \frac{1}{\theta}\sum_{i=1}^{n_\bd}d_i^2+n_\bd\ln\left[2\pi\theta\right].\label{e:-2lnL1}
    \end{align}%
    From here it can be seen that a single number, the sum of the square of the data
    \begin{align}
        x &= \sum_{i=1}^{n_\bd} d_i^2,\label{e:sufficient}
    \end{align}
    is a minimal sufficient statistic.
    Maximising the (logarithm of the) likelihood with respect to the variance relates the value of the statistic to the variance 
    \begin{align}
        \frac{\partial\ln\mathcal{L}\(x|\theta\)}{\partial\theta} &= \frac{x}{2\theta^2}-\frac{n_\bd}{2\theta}\nonumber\\
        &=0
	\end{align}
	so that
	\begin{align}
        x &= n_\bd\theta.
    \end{align}
    The Fisher information is calculated using equation~\eqref{e:fisher}
    \begin{align}
        \bF &=\frac{x}{\(\theta^{\rm fid}\)^3}-\frac{n_\bd}{2\(\theta^{\rm fid}\)^2}\nonumber\\
        &= \frac{n_\bd}{2\(\theta^{\rm fid}\)^2}.\label{e:analytic_fisher}
    \end{align}
    For $n_\bd=10$ and a fiducial variance of $\theta^{\rm fid}=1$ the Fisher information is
    \begin{align}
        \bF &=5.\label{e:fisheranalytic}
    \end{align}
    Since the single summary is a non-linear combination (squared sum) of the data, linear combinations will not be able to provide a single sufficient statistic.

\begin{figure}
        \centering
        \includegraphics{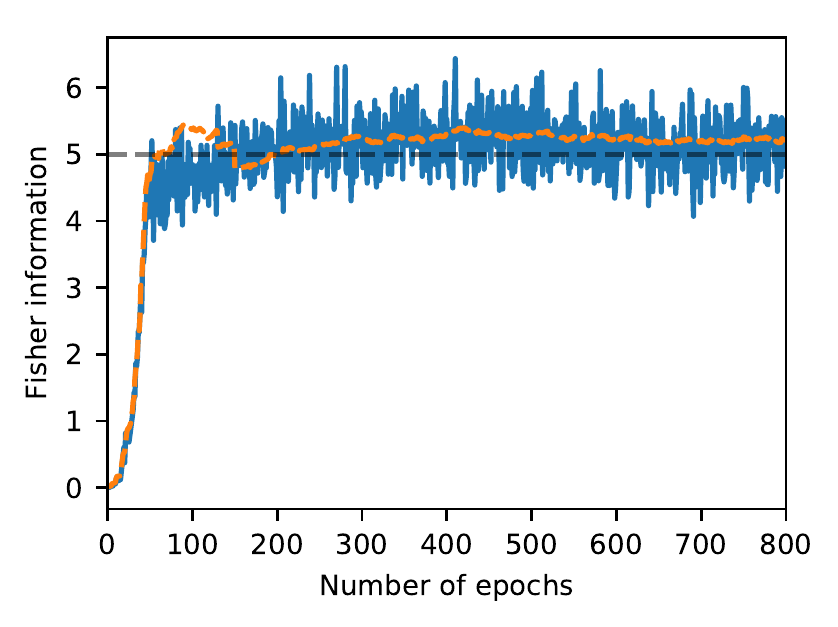}
        \caption{Value of the Fisher information obtained by the network at the end of each epoch of training.
        The solid blue line shows the Fisher information obtained by running a set of 500 simulations (and 50 partial derivatives), which are contained in the training set, through the network.
        The dashed orange line shows the Fisher information obtained by running the same number of simulations through the network, but where none of the simulations are present in the training set.
        The maximum amount of Fisher information expected is $\bF = 5$, show as a black dashed line.
        It is clear that the network manages to extract the entirety of the information given the data.}
        \label{fig:training}
    \end{figure}

\medskip
    Now consider training a network to maximise the Fisher information whilst summarising the data, as laid out in section~\ref{sec:FAAN}.
    We show the progress an example network makes until it extracts the full information in figure~\ref{fig:training}.

\medskip
	The fully connected network has two hidden layers with 256 neurons in each.
	We denote this configuration $[256, 256]$. The network uses leaky {\ttfamily ReLU} activation with $\alpha=0.01$ and a learning rate of $\eta = 0.01$.
    Each of the weights, $\bw^l$, are initialised with a value drawn from a normal distribution with mean $\mu = 0$ and standard deviation $\sigma = \sqrt{2/\kappa^{l-1}}$ where $\kappa^l$ is the number of neurons in layer $l$~\citep{He:2015}.
    As is usual when using the {\ttfamily ReLU} family of activation functions, the biases $\bb^l$ are initialised with a slightly positive value~\citep{Glorot:2010}, where $\bb^l = 0.1$ has been chosen here.
    To mimic the small number of simulations which would be available for complex data sets, we limit the total number of simulations to 1000 (+ 100 simulations created above and below the fiducial parameter value to calculate the numerical derivatives).
    These are divided into $n_{\rm train}=2$ training batches per epoch, such that $n_\bs = 500$ and $n_{\partial\theta}=50$.
    The training batches are split to provide variation in the statistical quantities $\bmu_\mf,_\alpha$ and $\bC_\mf$ when jumbling the simulations at the beginning of each epoch of training.
    We train the network for 800 epochs.
    To prevent overfitting, where the network learns features in the training set which are not present in the test data, 50\% of the neurons are dropped from the network on each batch of training.
    
\medskip
    From equation~\eqref{e:fisheranalytic}, it can be seen that the maximum Fisher information attainable for this problem is $\bF=5$.
    Figure~\ref{fig:training} shows that $\bF=5.15\pm0.39$ is obtained by the network over the last 10\% of the training epochs.
    The solid blue line in figure~\ref{fig:training} is the value of the Fisher information obtained from the network summaries of $n_\bs=500$ and $n_{\partial\theta}=50$ simulations from the training set (a single batch with no dropout), whilst the dashed orange line is the same for simulations which are not contained in the training set.
    We find that we are able to obtain a Fisher information slightly above $\bF=5$ as indicated by the straight black dashed line.
    This is because the data sets fluctuated to have a smaller variance than $\theta=1$ and therefore the Fisher information for these sets is higher than their expectation.
    The network interprets the fluctuation in the data as an indication that more information about the parameters is available from the network than is truly available.
    
\medskip
    We have found that a very large variety of hyperparameters will provide us with approximately $\bF=5$.
    Most notably we can use very deep networks with few neurons such as [5, 5, 5, 5, 5] to extremely simple networks with large numbers of neurons, i.e. [2048, 2048], each with very similar outcomes.
    The main difference with different architectures, that we have found, is the number of epochs necessary to maximise the Fisher information matrix.
    We have chosen a simple network of [256, 256] since it seems to converge more quickly than other networks.
    
\begin{figure}
        \centering
        \includegraphics{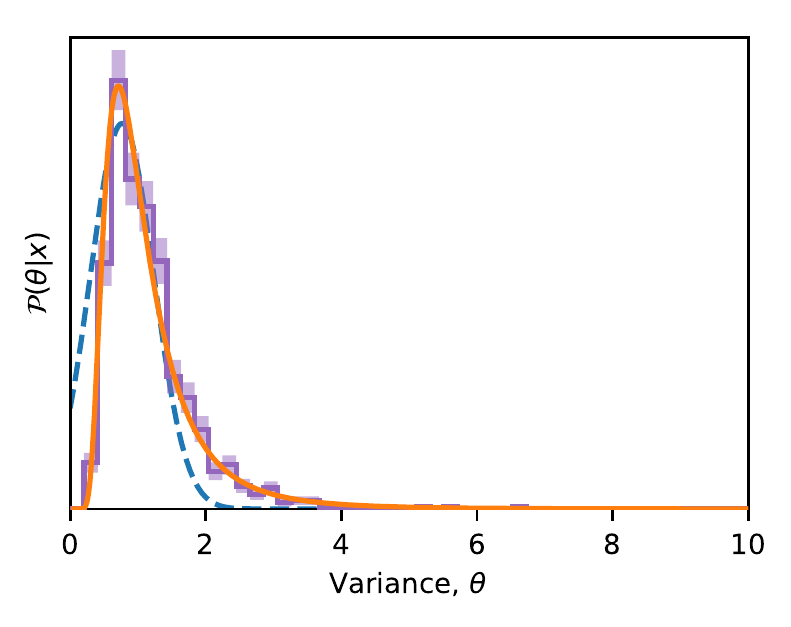}
        \caption{The posterior distribution for the variance of the real data.
        The solid orange curve is the analytic posterior distribution using Bayes' theorem and the likelihood in equation~\eqref{e:likelihood}.
        The dashed blue curve shows the posterior calculated from the asymptotic likelihood from the network summary and in purple is the ABC posterior obtained through PMC with the purple shaded error bars showing the 1-$\sigma$ Poisson width.
        Each distribution is normalised such that its integral is unity in the interval $\theta = [0, 10]$.
        We can see that the analytic posterior in the solid orange curve overlaps the PMC-ABC posterior in the purple histogram showing that the network has successfully learned to summarise the data.
        The blue dashed curve peaks at the same place as the analytic posterior with a similar width, which shows that the first order approximation of the posterior is also correct.}
        \label{fig:posterior}
    \end{figure}

\medskip
    Since the test model can be written down analytically, the true posterior distribution for some simulated test data $\bd$ (shown in table~\ref{tab:data}) can be found and is plotted as the solid orange curve in figure~\ref{fig:posterior}.
    The prior distribution used here is uniform between $\theta=(0, 10]$.
    \begin{table}
        \center
        \vskip1em
        \begin{tabular}{ll}
            \hline\hline
            Data & \hskip2.5em Value\\\hline  
            $d_1$&$-0.91903399$\\
            $d_2$&$-0.37322515$\\
            $d_3$&$-0.05613342$\\
            $d_4$&$\phantom{-}1.20816746$\\
            $d_5$&$\phantom{-}0.07649269$\\
            $d_6$&$-0.47171141$\\
            $d_7$&$-1.4756571$\\
            $d_8$&$-0.62946463$\\
            $d_9$&$-1.30334079$\\
            $d_{10}$&$-0.41441639$
            \\\hline\hline
        \end{tabular}
        \caption{Values of the input parameters for the original simulated true data set where the data is Gaussian noise $\bd = \left\{d_i \curvearrowleft \mathcal{N}\(0, 1\)\big|\,i \in[1,\, n_{\bd}]\right\}$.}
        \label{tab:data}
    \end{table}
    A first approximation of the posterior distribution using the network, without creating any additional simulations can be found using the asymptotic likelihood by expanding equation~\eqref{e:fgaussian} about the fiducial variance with $\Delta\theta=(-1, 9]$\footnote{This approximation is only true for ${\rm Abs}[\Delta\theta]\ll1$. 
    The interval chosen here is used only for plotting purposes.}.
    The asymptotic likelihood result is plotted in figure~\ref{fig:posterior} in dashed blue.
    It can be seen that the peak of the posterior found using the asymptotic likelihood corresponds with the peak of the analytic posterior, although as expected the rest of the distribution quickly deviates from the analytic result.
    To perform PMC-ABC, $N=1000$ parameter values, $\{\theta_k^{\,0}|\,k\in [1,\, N]\}$, are drawn from the uniform prior distribution, $p(\btheta)$, between $\theta=(0, 10]$.
    The PMC procedure, described above, is then carried out to obtain 1000 samples from the approximate posterior.
    Using a criterion that there needs to be 2000 draws of $\theta_k^{\,t}$ in iteration $t$ to be convinced that the approximate posterior has converged requires a total of 10232 simulations.
    The width of the acceptance parameter is $\varepsilon^T=0.086$ at the last iteration, $T$, meaning that the network summary of each of the accepted network summaries are within a band of $x^{\bs\,T}_{ik} = x\pm0.086$ of the network summary of the real data, $x$. 
    The histogram of the accepted points are shown in figure~\ref{fig:posterior} in purple.
    The PMC-ABC posterior distribution follows the analytic posterior distribution exactly, showing that the network has successfully learned how to summarise the data.

\medskip
    \begin{figure}
        \centering
        \includegraphics{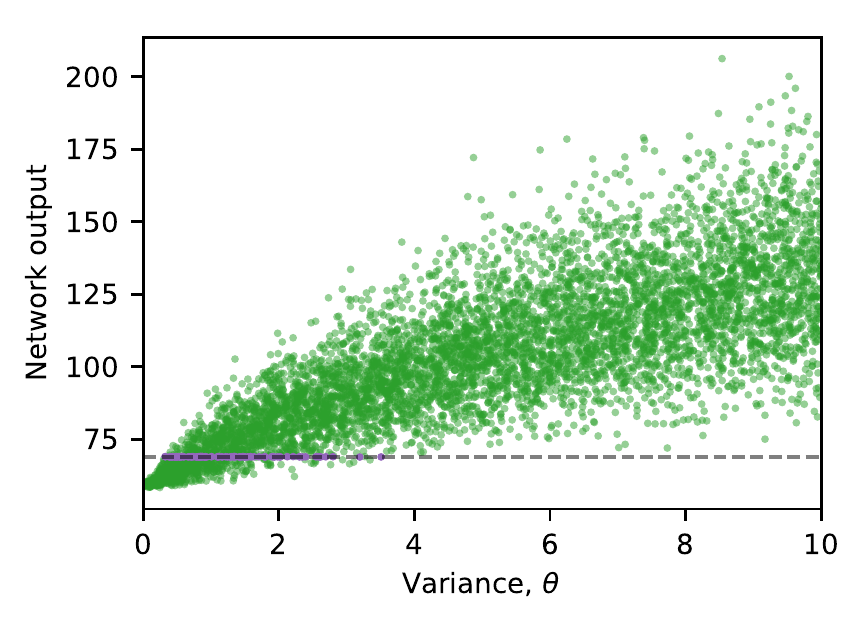}
        \caption{Network output as a function of the variance used to create the simulations.
        The green points are the network summaries of a selection of the simulations created from random draws from $\theta = (0, 10]$ for the random ABC procedure.
        The purple points are the accepted network summaries of the 1000 simulations within $x^{\bs\,T}_{ik} = x\pm\varepsilon^T$ with $\varepsilon^T=0.086$.
        The black dotted line indicates the network output of the real data.
        There is a strong correlation between the network output and the value of $\theta$ which suggests that the network has learned how to summarise the network input with respect to the model parameters.}
        \label{fig:network}
    \end{figure}%
    It is interesting to see the network outputs as a function of $\theta$, without using the PMC procedure.
    By performing ABC by randomly drawing from the whole prior, and not honing in on the true distribution, we can plot the network output as a function of the variance drawn from the prior, shown in figure~\ref{fig:network}. 
    The green points show the rejected samples and the purple points (under the black dashed line) show the accepted draws.
    The black dashed line shows the network output of the real data.
    There is a strong correlation between the network summary of the simulations and the value of $\theta$ used to create the simulation.
    Requiring that there are 1000 samples whose summaries are within $x^{\bs\,T}_{ik} = x\pm\varepsilon^T$, where $\varepsilon^T=0.086$, necessitates more than 600,000 draws from the prior, 50 times more draws than the PMC needs.
     It should be noted that the network summary is not equal to the value of $\theta$ and, in general can vary a lot by changing the network architecture, the initialisation of the weights or even just changing the order of the simulations used to train the network.
     The variation in the network summary is a manifestation of how the Fisher information is invariant under linear scalings of a sufficient statistic, although the scale of the statistic is able to be constrained somewhat by coupling the Fisher information matrix to the covariance of the outputs, as in equation~\eqref{e:realloss}.

\begin{figure}
        \centering
        \includegraphics{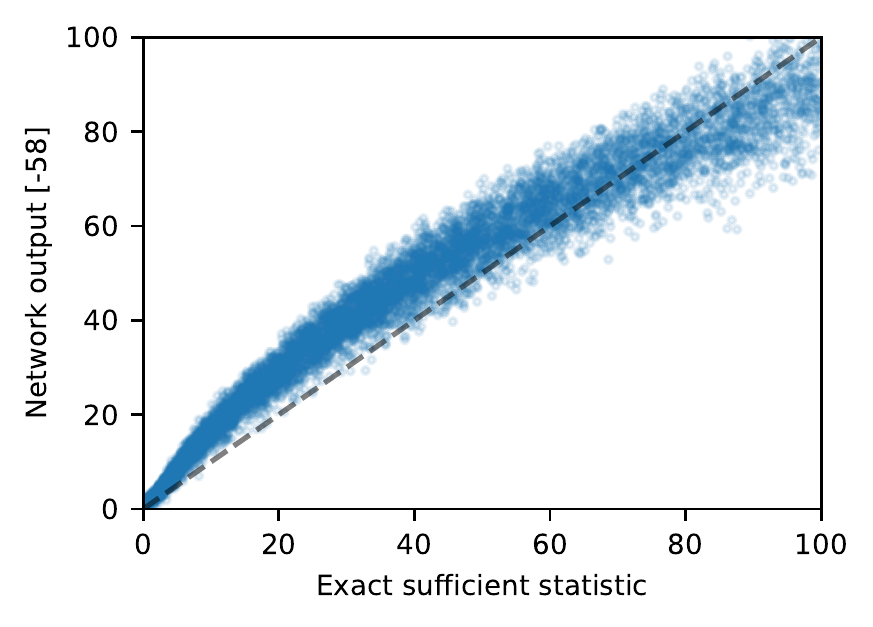}
        \caption{Rescaled network output for a given exact sufficient statistic.
        The blue dots show a scaled value of the network output at the exact sufficient statistic of simulations obtained at a range of $\theta$ during ABC.
        The black dashed line shows the expected value of the network output if the network had learned the map from data to the sufficient statistic perfectly.
        Since the scatter of the exact sufficient statistic to the network output closely follows the black dashed line, we know the network has approximately learned the correct map from data to sufficient statistic.
        There is a slight curve which arises from the fact that any one-to-one function of the sufficient statistic is still a sufficient statistic and so is of no concern.
        There is also a superficial broadening of the curve which shows that the map is only approximately correct.
        }
        \label{fig:nework_summary}
    \end{figure}

\medskip
     When creating simulations during the ABC procedure we can calculate the true sufficient statistic, i.e.
    \begin{align}
    	x^\bs_i & = \sum_{j=1}^{n_\bd}\left(d^\bs_{ij}\right)^2
    \end{align}
    where $i$ labels the random initialisation of the simulation and the the j labels the data point in the data set $\bd$.
    Plotting the exact sufficient statistic against the network output allows us to see how well the correct function is learned by maximising the IMNN, as seen in figure~\ref{fig:nework_summary}.
	The blue points show the values of exact sufficient statistics of the simulations and scaled values of the network outputs of the same simulations.
    The network output must be scaled due to the allowed linear scaling of the sufficient statistic.
    We actually found that network output is approximately 
    \begin{align}
    	{\rm network\ output} & \approx \sum_{j=1}^{n_\bd} \left(d_{ij}^\bs\right)^2 + 58,
    \end{align}
    without a linear scaling of the exact sufficient statistic, but with an offset.
    The black dashed line shows what would be expected if the exact map was learned by the network.
    We can see that the network output generally follows the sum of the square of the data closely with hints of a slight bend and superficial broadening at larger exact sufficient statistics.
    The bending is of no concern since any one-to-one function of the sufficient statistic is still a sufficient statistic, and we can see that the network output is clearly a monotonic function of the real summary.
    The broadening indicates that only an \emph{approximate} map is learned because the training of the network is incomplete due to lack of diversity within simulations and perhaps a sub-optimal choice of network hyperparameters.
    With greater variety within the simulations or, likewise, a greater number of simulations, the optimal map could be learned even more precisely.
    Nevertheless, we can see how minor an effect the broadening of the exact sufficient statistic is by looking at the results in figure~\ref{fig:posterior}.
    The resulting posterior distribution is equivalent to the analytic posterior, which is the real proof that the network has found the correct summary statistic.
           
    \subsubsection{Summarising Gaussian signals with known noise variance\label{ssec:known}}
    
    Now consider some noisy data where the real data $\bd = \big\{d_i \curvearrowleft \mathcal{N}\big(0, \theta + \sigma_{\rm noise}^2\big)\big|\,i \in [1,\, n_{\bf d}]\big\}$ has a signal variance of $\theta^{\rm true}=1$ and the variance of the noise is taken to be known $\sigma^2_{\rm noise}=1$.
    Simulations of the noisy data can be created and used to train the network, as before.
    The addition of the noise makes the likelihood less peaked about the \emph{true} parameter value and so the Fisher information is expected to be less than in original problem.
    Since the likelihood is known analytically, using equation~\eqref{e:analytic_fisher} it can be seen that $\bF=1.25$.
    The network manages to achieve $\bF \approx 1.25$ by the end of training, suggesting the network is capable of extracting close to the maximum amount of information possible.
    We have used a slightly less complex network here with [128, 128], but all other parameters the same.
    Again, many different architectures work equally well, but do not necessarily converge as quickly.
    We train the network for 2000 epochs before the Fisher information saturates to its maximum value.
    
    \medskip
    \begin{figure}
        \centering
        \includegraphics{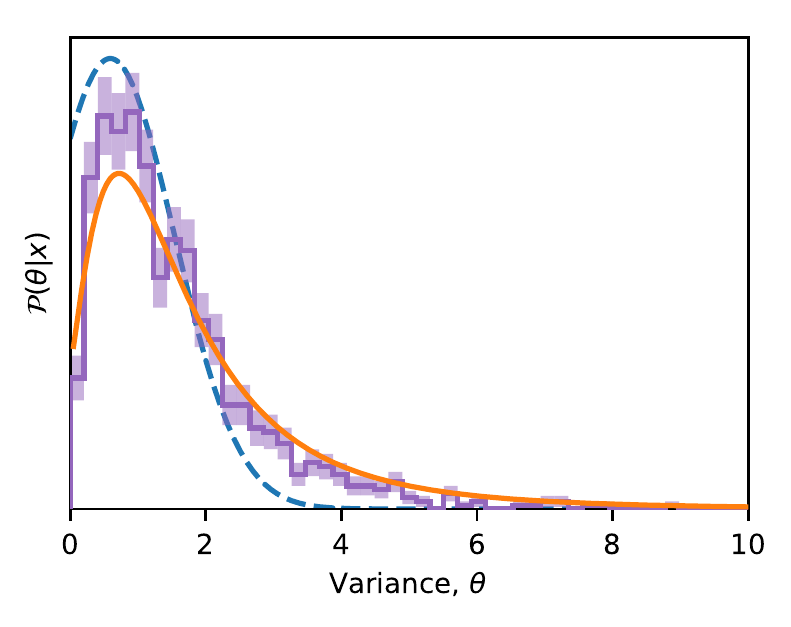}
        \caption{The posterior distribution of the signal variance of the Gaussian noise when the data is contaminated with known noise of $\sigma_{\rm noise}^2 =1$. 
        The solid orange line shows the analytic posterior distribution, whilst the posterior distribution from the asymptotic likelihood is shown in dashed blue and the purple histogram with 1-$\sigma$ Poisson shaded error bars shows the approximate posterior distribution from PMC-ABC.
        Each distribution is normalised such that its integral is unity in the interval $\theta = [0, 10]$.
        We can see that the solid orange curve and the purple histogram overlap along the entire range of $\theta$ suggesting that the network has learned the correct way to summarise the data.}
        \label{fig:posterior_known_noise}
    \end{figure}
    In figure~\ref{fig:posterior_known_noise}, it can be seen that the PMC-ABC posterior distribution, shown in the purple histogram with shaded 1-$\sigma$ Poisson regions, when given some simulated test data, $\bd$, is very similar to the analytic result shown by the solid orange line.
    The dashed blue approximate posterior distribution from the asymptotic likelihood again peaks very close to the maximum of the analytic posterior.
    The posterior distribution becomes maximal at the most likely parameter value given the data, with the variance given by the inverse Fisher information at the end of training.
	There are 1000 samples used to create the histogram of the PMC-ABC posterior which required approximately $2\times10^5$ simulations to be created during the PMC, where all samples are within $x^{\bs\,T}_{ik}=x\pm\varepsilon^T$, with $\varepsilon^T=0.109$.
    Since the analytic posterior distribution is so similar to the PMC-ABC posterior we can see that, even though the network is only given noisy simulations, it is capable of finding the true function to summarise the data.
    
    \subsubsection{Summarising Gaussian signals with unknown noise variance\label{ssec:unknown}}
    
    Now consider the problem where, again, the real data $\bd = \big\{d_i \curvearrowleft \mathcal{N}\big(0, \theta + \sigma_{\rm noise}^2\big)\big|\,i \in [1,\, n_{\bd}]\big\}$ has a signal variance of $\theta=1$ and the variance of the noise is also unknown with a uniform prior $\sigma^2_{\rm noise} \in (0, 2]$.
    \begin{figure}
        \centering
        \includegraphics{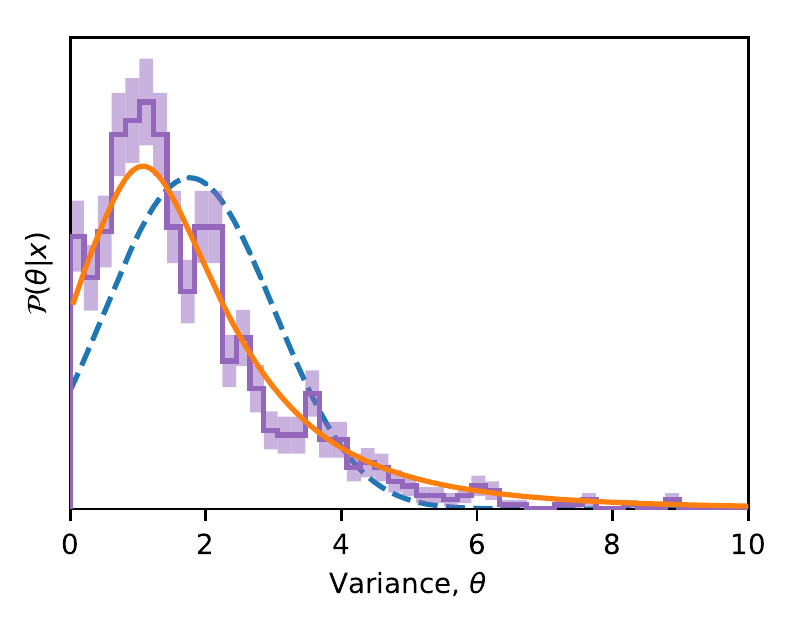}
        \caption{The posterior distribution of the signal variance when the data is contaminated with unknown noise $\sigma^2_{\rm noise} = (0, 2]$.
        The exact posterior is shown by the solid orange curve, the posterior distribution obtained using the asymptotic likelihood is in dashed blue and the PMC-ABC posterior with samples drawn using PMC is indicated by the purple histogram with shaded 1-$\sigma$ Poisson error bars.
        Each distribution is normalised such that its integral is unity in the interval $\theta = [0, 10]$.
        Even with unknown noise the network can summarise the data equally as well as a Rao-Blackwell estimate of the analytic case, leading to equivalent posterior distributions.
        The posterior distribution obtained from the asymptotic likelihood does not agree with the other distributions since the training simulations are not representative of the real data.
        }
        \label{fig:posterior_unknown_noise}
    \end{figure}
    We train using 1000 simulations (+100 for each of the derivatives) at a fiducial $\theta^{\rm fid}=1$ each with a different $\sigma^2_{\rm noise}$ randomly drawn from the uniform prior on the noise.
    The final value of the Fisher information from the network is less than in either of the two previous cases at $\bF=0.9$ using a slightly more complex network than in the previous section with an architecture of [128, 128, 64] but all other parameters the same.
    If the noise were assumed to be known at $\sigma^2_{\rm noise}=2$ then the maximum Fisher available, as calculated from equation~\eqref{e:analytic_fisher} would be $\bF=5/9$.
    The posterior distributions for $\theta$ are shown in figure~\ref{fig:posterior_unknown_noise}.
    Since the noise is unknown, a Rao-Blackwell estimate of the analytic distribution is made.
    Here, the posterior distribution is calculated for a range of given noise values from $\sigma_{\rm noise}^2=(0, 2]$ and their results summed at each value of $\theta$, plotted with a solid orange line.
    The PMC-ABC posterior is given by the purple histogram consisting of 1000 samples, which required approximately $10^5$ simulations using the PMC.
    Again, as before, the constraints on $\theta$ are incredibly similar to the analytic result, confirming that the network can approximate the exact summary very well.
    The Rao-Blackwell estimation procedure is also carried out to obtain the posterior calculated from the asymptotic likelihood, in dashed blue, although the result does not agree with the exact or PMC-ABC posteriors.
    The lack of agreement arises because the simulated test data is not well represented in the training simulations.
    Even though there is an under representation in the data, the network has learned the correct way to summarise data independent of the input, i.e. the network calculates the sum of the square of the input.
    
    \subsection{Summarising Gaussian signals with wrong fiducial variance\label{ss:fid}}
    
\begin{figure}
        \centering
        \includegraphics{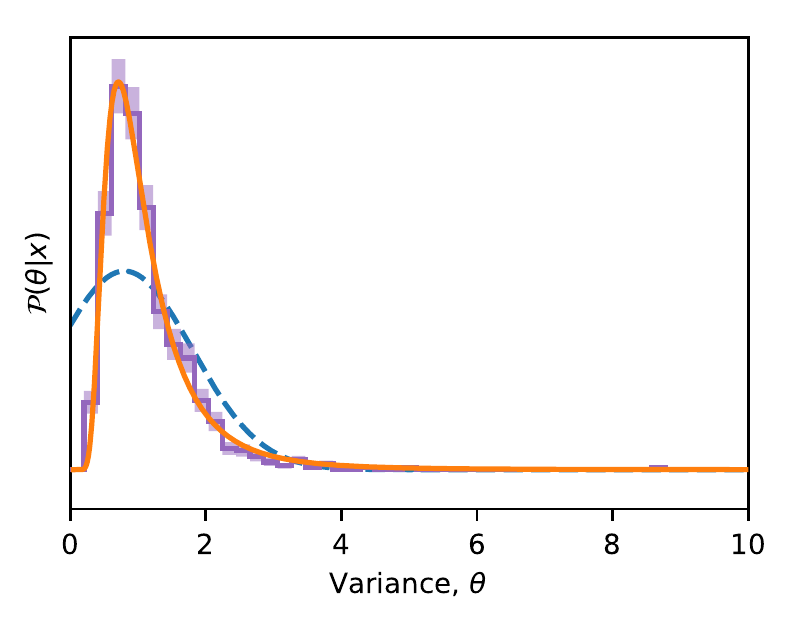}
        \caption{The posterior distribution of parameter $\theta$ where the real data is that of table~\ref{tab:data}, but the network has been trained with a fiducial $\theta^{\rm fid}\ne\theta^{\rm true}$. 
        The solid orange line shows the analytic posterior distribution and the purple histogram, with shaded 1-$\sigma$ Poisson widths, shows the approximate posterior distribution from PMC-ABC.
        The dashed blue curve shows the posterior distribution from the asymptotic likelihood.
        Each distribution is normalised such that its integral is unity in the interval $\theta = [0, 10]$.
        The analytic posterior distribution and the PMC-ABC posterior are again identical, which shows how the network is able to find the correct function to map data to summaries, even when the fiducial training parameter value is incorrect.}
        \label{fig:posterior_with_wrong_fiducial_training}
    \end{figure}

    Since the network trained in the known noise problem, in section~\ref{ssec:known}, is akin to a network trained at a fiducial parameter $\theta^{\rm fid}=2$, we can use it to test how well the network can predict the variance when the fiducial value does not coincide with the true parameter.
    It would be expected that data with $\theta = 1$ would be under-represented in a training data set where the fiducial value is $\theta^{\rm fid} = 2$.
    Na\"ively, one would assume that the network would not perform as well as a network trained using simulations created at $\theta^{\rm fid} = 1$, especially since the Fisher information available from this network is $\bF=1.25$ and not $\bF=5$ as in section~\ref{ss:gaussian}.
    However, figure~\ref{fig:posterior_with_wrong_fiducial_training} shows that the parameter constraints given the same real data as in table~\ref{tab:data} are equally as strong as when using the trained network from section~\ref{ss:gaussian}.
    It is promising that the training of the network seems fairly insensitive to the choice of fiducial parameter. 
    The posterior distribution from the asymptotic likelihood, in dashed blue, is much wider than the same curve in figure~\ref{fig:posterior} since the variance of the distribution is given by the Cram\'er-Rao bound, i.e. $\bF^{-1}=0.8$, rather than $\bF^{-1}=0.2$ when the network from section~\ref{ssec:known}.
    The fact that the purple histogram matches the analytic solid orange distribution so well indicates that the network has learned the correct way to summarise data, rather than learning an algorithm for mapping simulations to an output which specifically depends on the fiducial parameter value.
    For example, in the problem considered here, we know that the correct summary of the data is the sum of the square of the data (or at least a linear scaling of the sum of the square of the data).
    The network is trained in such a way that the abstract function of weights, biases and inputs that the network represents closely approximates the sum of the square of the input.
    Once abstract function is learned, it does not matter what parameter value is used to create the simulations, even if that parameter is far from the fiducial value, because the network will still output the sum of the square of the input.
    It is extremely encouraging to see that the network can extrapolate beyond its training data by depending on the robustness of the learned patterns.
    
    \subsection{Summarising quasar spectra}\label{ss:lyman}
    
    Beyond the elementary test case on variance estimation, we can consider models that are of more astronomical interest.
    Here we attempt to generate constraints on the amplitude of scalar perturbations, $A_{\rm s}$, using a simplistic 1D model of the Lyman-$\alpha$ forest from a single quasar.
    To generate simulations we begin by using the halo mass function calculator {\ttfamily hmf} module~\citep{Murray:2013qza} in {\ttfamily python} to generate the 3D power spectrum $P^{\rm 3D}(k)$, evolved using the method of Eisenstein and Hu~\citep{Eisenstein:1997ik}, at a redshift of $z=2.25$ with fixed cosmological
    parameters (at $z=0$).
    The cosmological parameters come from the Planck 2015 temperature and low-$\ell$ polarisation results~\citep{Ade:2015xua}, $H_0=67.7$ km Mpc$^{-1}$s$^{-1}$, $\Omega_{\rm m}=0.307$, $\Omega_{\rm b}=0.0486$, $n_{\rm s}=0.9667$, $\sigma_8=0.8159$, $T_{\rm CMB}=2.725$ K, $N_{\rm eff}=3.05$, and $\sum m_\nu=0.06$ eV, calculated using {\ttfamily astropy}~\citep{Astropy:2013}.
	The power spectrum is calculated between $\ln k_{\rm min}/(1h{\rm Mpc}^{-1})=-18.42$ and $\ln k_{\rm max}/(1h{\rm Mpc}^{-1})=9.90$ in steps of $\Delta\ln k/(1h{\rm Mpc}^{-1}) = 0.005$.
    The correlation function can be found using
    \begin{align}
    	    \xi(r) & = \int_0^\infty \frac{dk}{2\pi^2}\exp\left[-R_w^2 k^2\right]k^2P^{\rm 3D}(k){\rm sinc}(kr)    
    	\end{align}
    where the exponential term is a smoothing function where we use $R_w = 5h^{-1}{\rm Mpc}$.
    We calculate the value of $\xi(r)$ between $-200 < r < 200h^{-1}{\rm Mpc}$ in $N = 8192$ bins.
    To simulate the density fluctuations along the line of sight, we calculate the 1D power spectrum using
    \begin{align}
    	P^{\rm 1D}(k) & = \int_{-\infty}^\infty dr\exp[ikr]\xi(r).
    \end{align}
    The Lyman-$\alpha$ peak in the rest frame of an emitter is $\lambda^\alpha_{\rm RF}=121.567{\rm nm}$~\citep{Bautista:2015} and we use the fact that BOSS can measure absorbers in the redshift range $1.96<z<3.44$~\citep{Bautista:2017zgn}.
    Using
    \begin{align}
        z & = \frac{\lambda}{\lambda_{\rm RF}} - 1\label{e:redshift}
    \end{align}
    the minimum observed wavelength of the Lyman-$\alpha$ peak is $\lambda_{\rm min}=359.838\,{\rm nm}$ (at $z = 1.96$) and the maximum wavelength is $\lambda_{\rm max}=539.757\,{\rm nm}$ (at $z = 3.44$)~\citep{Bautista:2017zgn}.
    The length $L$ of the survey in comoving space is calculated between these redshifts, yielding $L = 1122.9 h^{-1}\,{\rm Mpc}$. 
    The frequency spacing is given by the inverse of the survey length, so we consider a range of $k = (0, 14.6]h\,{\rm Mpc}^{-1}$ with $N = 8192$ bins.
    We modify the 1D power spectrum such that it more closely follows the gas power spectrum as seen from Lyman-$\alpha$ absorptions~\citep{McDonald:2003},
    \begin{align}
        P_{\rm g}^{\rm 1D}(k) & = \beta D(k,\mu)P^{\rm 1D}(k)\label{e:powerspectrum}
    \end{align}
    where $\beta$ is a free parameter, set for a given realisation of noise which ensures that $\langle F\rangle=0.8$~\citep{Font-Ribera:2012}.
    $D(k, \mu)$ is a term which modifies the small-scale power spectrum~\citep{McDonald:2003} and is of the form
    \begin{align}
        D(k,\mu)  & = \exp\left[\left(\frac{k}{k_{\rm NL}}\right)^{\alpha_{\rm NL}} - \left(\frac{k}{k_{\rm P}}\right)^{\alpha_{\rm P}} - \left(\frac{k_\parallel}{k_{\rm V}}\right)^{\alpha_{\rm V}}\right],
    \end{align}
    where
    \begin{align}
        k_{\rm V} & = k_{{\rm V}_0} \left(1 + \frac{k}{k'_{\rm V}}\right)^{\alpha'_{\rm V}}
    \end{align}
    and $k_{\rm NL} = 6.40\hMpc$, $\alpha_{\rm NL} = 0.569$, $k_{\rm P} = 15.3\hMpc$, $\alpha_{\rm P} = 2.01$, $k_{{\rm V}_0} = 1.220$, $k'_{\rm V} = 0.923\hMpc$, $\alpha'_{\rm V} = 0.451$, $\alpha_{\rm V} = 1.50$~\citep{Blomqvist:2015} and we choose to use $\mu=k_{||} / k = 1$  since we only consider independent quasar lines, i.e. the flux is completely decorrelated from one line to the next. 
    The above numbers are computed for the log-flux explicitly described in~\citep{McDonald:2003}.
    For the purpose of demonstration we keep the same $k$ dependence here.
    With the gas power spectrum in equation~\eqref{e:powerspectrum}, normalised by the length of the survey, we can generate 1D random Gaussian fields, $\delta_{\rm g}$.
    The Gaussian fields are generated by multiplying unit variance, zero mean Gaussian noise with $\(P^{\rm 1D}_{\rm g}(k)/2\)^{1/2}$ and Fourier transforming into real space, including the normalisation of $N / (2 L)$ due to the discrete nature and finite period of the discrete Fourier transform.
    The flux from quasars is absorbed by neutral hydrogen in over-densities in the density field, and can be calculated from the fluctuating Gunn-Peterson approximation~\citep{Peeples:2009uj} as
    \begin{align}
        F & = \exp\left[-\tau\right]
    \end{align}
    where we consider the form of the optical depth to be
    \begin{align}
        \tau & = 1.54 \left(\frac{T_0}{10^4{\rm K}}\right)^{-0.7}\frac{10^{-12}{\rm s}^{-1}}{\Gamma_{\rm UV}} \left(\frac{1 + z}{1 + 3}\right)^6 \\
        &\phantom{==}\times\frac{0.7}{h}\left(\frac{\Omega_{\rm b}h^2}{0.02156}\right)^2 \frac{4.0927}{H(z)/H_0} \rho^{2 - 0.7 \gamma}\label{e:gunnpeterson}
    \end{align}
    where $T_0=18400{\rm K}$ is the normalisation to the power-law temperature-density relation $T=T_0(1 + \delta_g)^{\gamma-1}$ with $\gamma = 0.29$ (both here and in equation~\eqref{e:gunnpeterson}) and $\Gamma_{\rm UV}= 4\times10^{-12}s^{-1}$ is the photo-ionisation rate due to the ambient UV background~\citep{Peeples:2009uj}. 
    The gas density field is normalised such that its mean is unity,
    \begin{align}
        \rho & = \frac{\exp\left[\delta_{\rm g}\right]}{\left\langle\exp\left[\delta_{\rm g}\right]\right\rangle}.
    \end{align}
    The continuum flux can be calculated between the Lyman-$\alpha$ and Lyman-$\beta$ peaks at $\lambda^\alpha_{\rm RF}=121.567\;{\rm nm}$ and $\lambda_{\rm RF}^\beta=102.572\;{\rm nm}$ using the PCA formulation of~\citep{Suzuki:2005}.
    The continuum flux in the rest frame of the emitter is calculated using
    \begin{align}
        r(\lambda)  &= \mu(\lambda) + \sum_i c_i(\lambda)\xi_i(\lambda)\label{e:continuum}
    \end{align}
    where $\mu(\lambda)$ is the mean flux over many quasars, $\xi_i(\lambda)$ are the $i^{\rm th}$ principal components and $c_i(\lambda)$ are the amplitudes of the principal components which we consider to be $c_i(\lambda) = 1$ for simplicity.
    The continuum can be transformed into the observer's wavelength space by assuming a redshift for the quasar and inverting equation~\eqref{e:redshift}.
    We choose the redshift of the simulated (and real) quasar to be $z=2.91$.
    The flux, which is currently in real space, is transformed into wavelength space by interpolating the comoving distance, $r$, along given redshift values, $z$, using the {\ttfamily hmf} {\ttfamily comoving\_distance(}$z${\ttfamily)} function and then using equation~\eqref{e:redshift}.
    The continuum modulated flux from the quasar is simply
    \begin{align}
        f(\lambda) & = F(\lambda)C(\lambda)
    \end{align}
    where $C(\lambda)$ is $r(\lambda)$ from equation~\eqref{e:continuum} in the rest frame of the observer~\citep{Bautista:2015}.
    \begin{figure}
        \centering
        \includegraphics{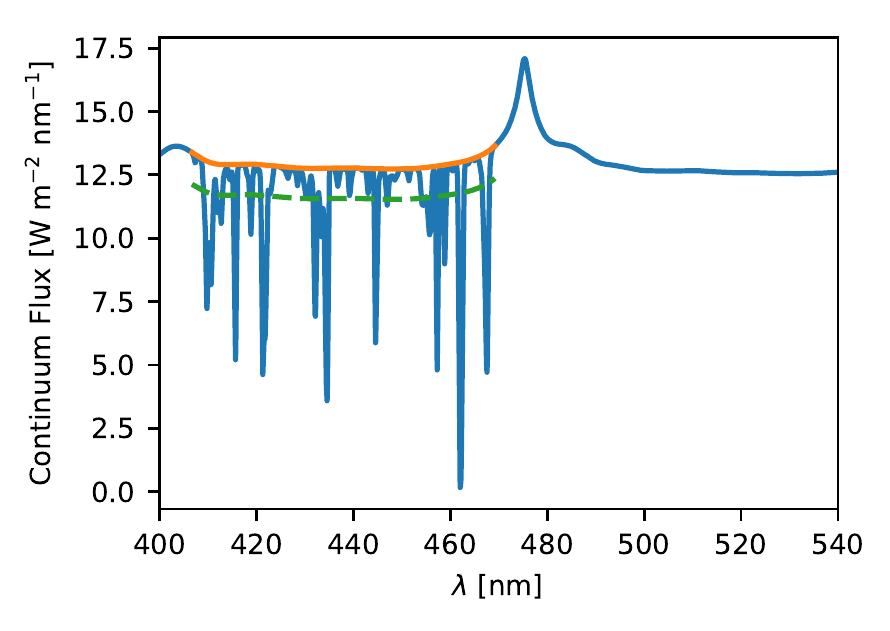}
        \caption{Simulated spectrum of a quasar at $z=2.91$ in blue, with the value of the continuum in orange (light) and the mean flux in the Lyman-$\alpha$ forest in dashed green.}
        \label{fig:continuum}
    \end{figure}
    Figure~\ref{fig:continuum} shows the generated flux from a single mock quasar at $z=2.91$ in blue.
    The orange (lighter) line shows the continuum flux between the Lyman-$\alpha$ and Lyman-$\beta$ peaks and the dashed green line shows the mean of the transmitted flux.
    We only consider the flux between $406.6{\rm nm}<\lambda<469.2{\rm nm}$ which is $104{\rm nm}<\lambda<120{\rm nm}$ in the rest frame of the quasar~\citep{Bautista:2017zgn}.
    We bin the wavelengths using the resolution from the BOSS coadded spectra of $\Delta\log_{10}\lambda/1{\rm nm} = 10^{-4}$~\citep{Bautista:2015} which gives a flux in ${\rm Wm}^{-2}{\rm nm}^{-1}$, but needs to be measured in photon counts.
    Using the method\footnote{In particular we use the method described in \url{http://www.sdss.org/dr12/algorithms/spectrophotometry/} in the section called ``DR9 Flux to Photons''. We use quasar 024918.47+025035.6 as a guideline.} in~\citep{Ahn:2012} we see that for a quasar such as the one we are generating the spectra for, there is an almost one-to-one correspondence between flux and photon count (albeit the photon count is integer)~\citep{Ahn:2012}.
    Therefore, we make the assumption that making the flux into integer values and then applying Poisson noise satisfactorily represents real quasar spectra.
    Our binned, noisy spectra have 581 data points, each of which can be used as an input to an IMNN.
    
\begin{figure}
        \centering
        \includegraphics{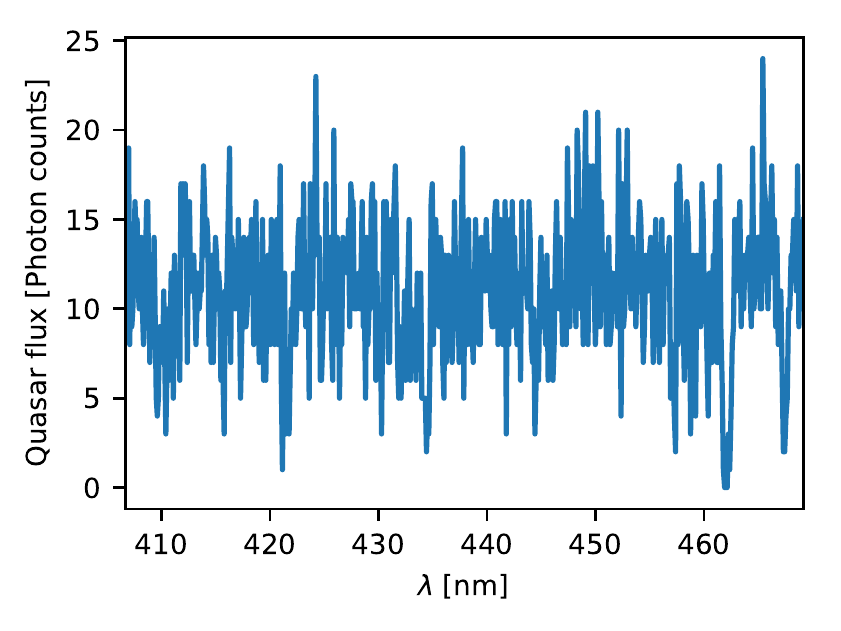}
        \caption{Simulated observation of a quasar spectrum from a quasar at $z=2.91$ between the Lyman-$\alpha$ and Lyman-$\beta$ peaks in the rest frame of the observer.
        We use this data as our simulated test data for the PMC-ABC.}
        \label{fig:lyman_input}
    \end{figure}

    An example of the simulated test data input to the network is shown in figure~\ref{fig:lyman_input}.

\medskip
    For any set of fixed cosmological parameters the value of amplitude of scalar perturbations, $A_{\rm s}$, is a scaling of $P^{\rm 1D}_{\rm g}(k)$. 
    To get constraints on $A_{\rm s}$ we can train a network at a fiducial $A_{\rm s}$ and then use the PMC to find the posterior distribution of $A_{\rm s}$ compared to some simulated test data.
    In fact, for simplicity, we can consider the parameter $\theta$ to be some multiplicative scaling of the amplitude, $A_{\rm s} = \theta A_{\rm cosmo}$ with $A_{\rm cosmo}$ the amplitude of the power spectrum found in equation~\eqref{e:powerspectrum}.
    We use $\theta^{\rm fid}=\exp[0]$ as the fiducial parameter, i.e. $A_{\rm s}^{\rm fid}=A_{\rm cosmo}$.

\medskip
    A relatively simple network, such as [256, 256], is able to obtain a Fisher information of $\bF=0.015$, which is the maximum Fisher information that could be found over a large range of different network architectures and hyperparameters.
    However, the network which was most resilient to incorrect fiducial values was more complex than those networks previously considered.
    The network with the largest Fisher information by the final epoch of training, which could handle incorrect fiducial parameters was a network four hidden layers shaped like $[1024, 512, 256, 128]$, using 1000 simulations (with 100 simulations each for the upper and lower components of the derivative) which were split into two batches, an initial bias of $\bb=0.1$, where the activation function is leaky {\ttfamily ReLu} with $\alpha=0.1$, a dropout of 20\% and a learning rate of $\eta=1\times10^{2}$ when training for 10000 epochs.

\begin{figure}
        \centering
        \includegraphics{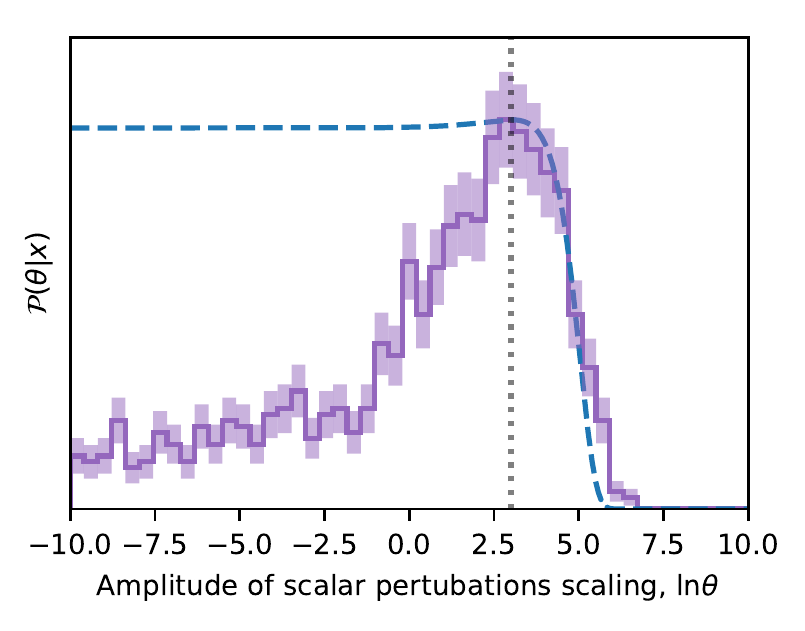}
        \caption{Posterior distribution for the scaling of the amplitude of scalar perturbations, $\ln\theta$.
        The dashed blue curve shows the Gaussian approximation to the true constraints as estimated with the training simulations and the purple histogram, with shaded 1-$\sigma$ Poisson error bars, is calculated from samples from PMC-ABC.
        The peak of both posterior distributions occur at the vertical black dotted line which shows the true value of the parameter.
        Although the constraints span several orders of magnitude, we expect these kinds of constraints from a single observation of a quasar absorption spectrum.
        Most importantly, we have shown that we can summarise extremely noisy data by solely maximising the Fisher information.}
        \label{fig:ly_posterior}
    \end{figure}

\medskip
    As before, once the network was trained, PMC-ABC could be performed.
    We used a uniform prior in logarithmic space of $\theta = \exp[-10, 10]$.
    The simulated test data was created away from the fiducial parameter value of $\theta^{\rm fid}= \exp[0]$ at $\theta^{\rm real} = \exp[3]$, i.e. $A_{\rm s} = \exp[3] A_{\rm cosmo}$, and is shown in figure~\ref{fig:lyman_input}.
    The posterior distribution for the value of $\theta$ can be found in figure~\ref{fig:ly_posterior}.
    Here, we required 1000 samples in the posterior requiring at least 2500 draws in the final iteration of the PMC to be convinced that the posterior had converged.
    The histogram peak, and the tentative peak of the leading order expansion of the likelihood, are at their maximum at $A_{\rm s} \approx \exp[3]A_{\rm cosmo}$, i.e. $\ln\theta\approx3$, which confirms that the correct test parameter can be recovered, shown as the vertical black dashed line in figure~\ref{fig:ly_posterior}.
    There is a large, degenerate tail in the PMC-ABC posterior which arises due to the amplitude of the random Gaussian noise, used to create the quasar spectrum, being so small that the features in the generated flux become negligible.
    Since the network output of the random fluctuations are still reasonably close to the network output from the real data, they cannot be constrained.
    The lack of constraining power at low $\theta$ is even clearer in the posterior from the leading order expansion.
    The constraints on $A_{\rm s}$ span approximately 5 orders of magnitude or more using the PMC-ABC posterior, which seems poor, but is due to using only one quasar spectrum to constrain cosmology with.
    Joint inference using several quasars would provide a much stronger constraint, as is done when using cosmological surveys.
    Although the constraints are not particularly strong, we have shown that we can learn to extract information from highly noisy data, and summarise it in such a way that we can perform PMC-ABC to get a posterior distribution for parameters of interest.
    
    \subsection{Gravitational waveform frequency\label{ss:graff}}
    
    \citep{Graff:2010dt} showed that the MOPED algorithm, described in section~\ref{sec:MOPED}, was unable to summarise the central oscillation frequency of a gravitational waveform from LISA without introducing spurious features~\citep{Graff:2010dt}.
    By using non-linear summaries of the data, the problem in~\citep{Graff:2010dt} can be avoided.

\medskip
    We start by considering a sine-Gaussian gravitational wave signal as could be seen in LISA, with a short burst duration and frequency space waveform~\citep{Feroz:2009eb} of
    \begin{align}
    	\overline{h}(f) & = \frac{AQ}{f}\exp\left[-\frac{Q^2}{2}\left(\frac{f - f_{\rm c}}{f_{\rm c}}\right)^2\right]\exp\left[2\pi i f_{\rm c}t_{\rm c}\right],
	\end{align}
	where $A$ is some amplitude, $Q$ is the width of the gravitational wave burst, $t_{\rm c}$ is the time of the burst and $f_{\rm c}$ is the central oscillation frequency.
	We fix $A=3.5$, $Q = 5$ and $t_{\rm c}=1\times10^5s$ and require that the signal-to-noise of the burst is $S/N=34$~\citep{Graff:2010dt}.
	We are interested in summarising and constraining the parameter $f_{\rm c}$.
    To generate a simulation of the gravitational wave signal, we use the one-sided noise power spectral density of the LISA detector~\citep{Feroz:2009eb}, which is
    \begin{align}
    	S_{h}(f) & = 16\sin^2\left[2\pi f t_{\rm L}\right] \bigg(2S_{\rm pn}\big(1 + \cos\left[2\pi f t_{\rm L}\right]\\
        & \phantom{=} \,\,\, + \cos^2\left[2\pi f t_{\rm L}\right]\big) + \big(\cos\left[2\pi f t_{\rm L}\right] / 2 + 1\big)S_{\rm sn} f^2\bigg),\nonumber\\
		S_{\rm pn}(f) & = \left(1 + \left(\frac{10^{-4}{\rm Hz}}{f}\right)^2\right)\frac{S_{\rm acc}}{f^2},
    \end{align}
    where $S_{\rm sn} = 1.8\times10^{-37}{\rm Hz}^{-1}$ is the shot noise, $S_{\rm acc} = 2.5\times10^{-48}{\rm Hz}^{-1}$ is the proof acceleration mass and $t_{\rm L}=16.678s$ is the light travel time along one arm of the LISA constellation.
    To generate the real space gravitational wave burst, we calculate the frequency space waveform $\overline{h}(f)$ and detector noise $\overline{n}(f)$ and then Fourier transform them into real space
    \begin{align}
    	\overline{h}(f) & = \int_{-\infty}^{\infty} dt\,h(t)\exp[2\pi ift],\\
    	\overline{n}(f) & = \int_{-\infty}^{\infty} dt\,n(t)\exp[2\pi ift].
    \end{align} 
    We perform the Fourier transform at 2048 time steps from $t=9.9 \times 10^4s$, sampled at $1s$ intervals.
    The output of the LISA detector is then given by 
    \begin{align}
		\bd & = \bh(\btheta^{\rm true}) + \bn
    \end{align}
    where $\bh(\btheta^{\rm true})$ is the values of the gravitational waveform at the true parameter values, $\btheta^{\rm true} = \{A^{\rm true}, Q^{\rm true}, t^{\rm true}_{\rm c}, f^{\rm true}_{\rm c}\}$ at the sampled time and $\bn$ is a random realisation of the noise.
    When assuming a noise covariance which is independent of the signal, $\sigma_{\rm n}^2 = \mathbb{I}$, the logarithm of the likelihood is particularly simple~\citep{Feroz:2009eb} and is given by
    \begin{align}
    	\ln\mathcal{L} & = C - \frac{||\bd - \bh(\btheta)||^2}{2},\label{e:LISAlikelihood}
    \end{align}
    where $\bh(\btheta)$ is the real space gravitational wave at, not-necessarily-true, parameters $\btheta$ and $C$ is a constant which we set to zero.
\begin{figure}
        \centering
        \includegraphics{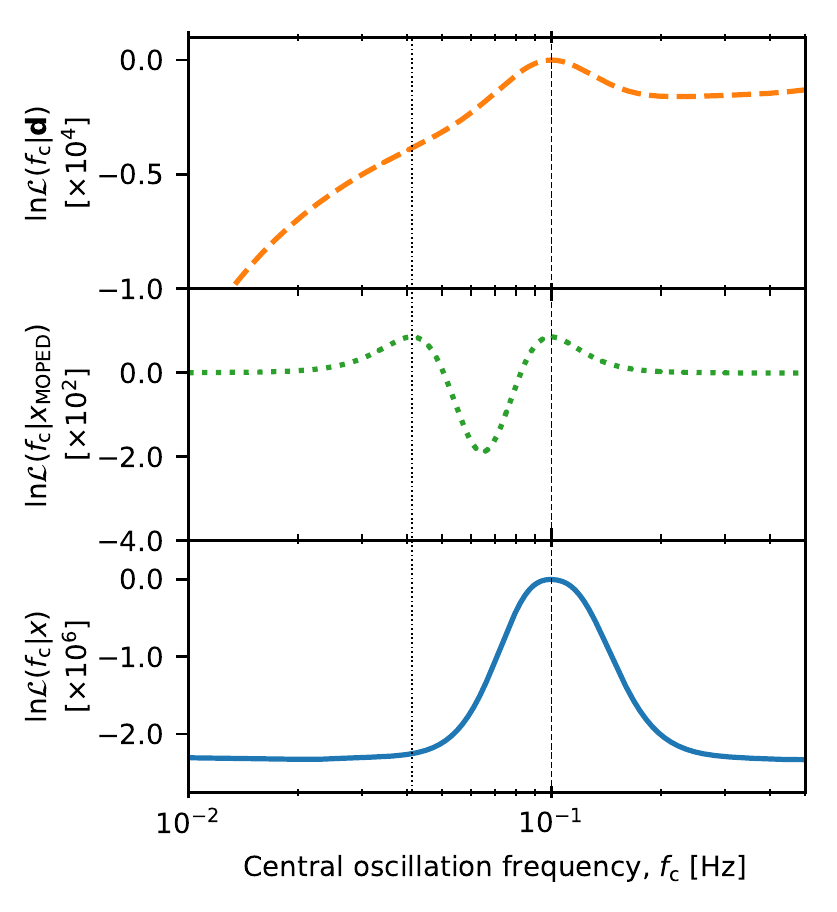}
        \caption{Logarithm of the likelihood for the central oscillation frequency, $f_{\rm c}$. 
        The dashed orange line in the upper panel shows the likelihood using all the data, whilst the green dotted line in the middle panel and the blue solid line in the bottom panel show the approximate Gaussian likelihoods when using compression, MOPED and IMNN respectively.
        All three likelihoods have peaks at the correct $f_{\rm c}^{\rm true}=0.1$Hz, but an aliasing peak arises in the MOPED likelihood due to a none-monotonic mapping from $\bh(f_{\rm c})\to x_{\rm MOPED}^{\bh}(f_{\rm c})$.
        The compression using the IMNN on the other hand does not have any aliasing peaks since the network has learned the non-linear map from data to frequency.}
        \label{fig:LISA}
    \end{figure}
    
\medskip
    We are interested in summarising the data to constrain the central oscillation frequency, $f_{\rm c}$, of the gravitational wave.
    To do so, we use a network which takes in the 2048 inputs from the data with the architecture [10, 10, 10, 10, 10].
    The network has a 10\% dropout and {\ttfamily leaky ReLU} activation with $\alpha = 0.01$.
    The learning rate is fixed at $\eta=10^{-5}$ and the biases are initialised slightly positively at $\bb^l=0.1$.
    We train for 1200 epochs using 1000 fiducial simulations and 100 simulations each for the positive and negative parts of the numerical derivative, all of which is split into two combinations.
    Once trained, we can use the network to summarise the data.
    We also use equation~\eqref{e:LISAlikelihood} to calculate the logarithm of the likelihood from the summary by passing the simulated test data, $\bd$, with a given realisation of the noise and generated at $f_{\rm c}^{\rm true}=0.1$Hz, through the network $\mf:\bd\to x$ and comparing it to the waveform at a given $f_{\rm c}$, $\mf:\bh(f_{\rm c})\to x^\bh(f_{\rm c})$.
    However, since the noise is included in the realisations which is passed through the network, the noise variance needs to be transformed as well.
    Assuming the variance is small, so that the likelihood remains Gaussian near the peak, the error propagation gives the new variance as
    \begin{align}
    	\sigma'^2_n & = \left|\frac{\partial x^{\bh}(f_{\rm c})}{\partial f_{\rm c}}\right|^2\sigma^2_n
	\end{align}	
    where the gradient should be evaluated at or near the true mean.
    The modified approximate likelihood, assuming Gaussian noise, for the IMNN summary evaluated at different parameters is therefore given by
	\begin{align}
    	\ln\mathcal{L} & = C - \frac{||x - x^\bh(f_{\rm c})||^2}{2\sigma'^2_n}.\label{e:LISAlikelihood_}
    \end{align}
    We calculate equations~\eqref{e:LISAlikelihood} and~\eqref{e:LISAlikelihood_} using simulated test data, $\bd$, generated at $f_{\rm c}^{\rm true}=0.1$Hz between $1\times10^{-2}<f_{\rm c}<0.5{\rm Hz}$.
    The logarithm of the likelihood of $f_{\rm c}$ calculated using all the data, $\ln\mathcal{L}(f_{\rm c}|\bd)$ is shown in the upper subplot of figure~\ref{fig:LISA} as a dashed orange line.
    This is compared to $\ln\mathcal{L}(f_{\rm c}|x_{\rm MOPED})$ using the MOPED summary as the dotted green line in the middle subplot and $\ln\mathcal{P}(f_{\rm c}|x)$ using the network summary as the solid blue line in the bottom subplot.
    The MOPED summary assumes a noise covariance which is independent of the signal such that the compression parameter is simply
    \begin{align}
    	{\bf r}_{f_{\rm c}} &\propto \bmu,_{f_{\rm c}}.
    \end{align}
    For the network summary, the noise is automatically included through the random initialisation of the simulations used to train the network.
    It can be seen that each of the likelihoods in figure~\ref{fig:LISA} agree with $f_{\rm c}^{\rm true}=0.1$Hz, shown with the dashed black line, but a false aliasing peak appears, shown with the dotted black line, when using the MOPED summary.
    This false maximum in the likelihood arises from unsuccessfully undoing the Fourier transform which leaves the mapping from $\bh(f_{\rm c})\to x^{\bh}_{\rm MOPED}(f_{\rm c})$ not being one-to-one.
    On the other hand, the IMNN compression does not suffer this problem.
   	There is a clear unique summary which, when used to calculate the approximate likelihood assuming Gaussian noise and evaluated at different $f_{\rm c}$, results in a single peak at the $f_{\rm c}^{\rm true}$.
    Full inference on $f_{\rm c}$ is then possible using PMC-ABC.
    \medskip
    
    This test shows that, through the use of the non-linear function provided by the IMNN, we are able to surpass the capability of linear compression.
    Not only can the summary from the network be at least as informative as the MOPED summary, it is also more robust since it is able to avoid misleading parameter inference due to non-trivial mappings.
    
    \section{Conclusions}
    
    We have shown how information maximising neural networks (IMNNs) can perform automatic physical inference.
    Automatic physical inference begins by training a neural network to find the optimal non-linear summaries of data supplied only with simulations and no other knowledge about how to best compress data.
    Once the network is trained, its output is used to perform PMC-ABC and find the approximate posterior distribution of any parameter that the network is sensitive to. 
    We have also shown that the network is insensitive to poor choice in fiducial parameter value when generating simulations.

\medskip
	We consider the technique presented in this paper as an extension or replacement to other massive optimal data compression procedures.
    The MOPED algorithm is able to optimally compress data using linear combinations under the assumption that the likelihood is known and is, to first order, Gaussian.
    Further, the method in~\citep{Alsing:2017var} generalises MOPED to any given likelihood function, where the compressed statistics no longer need to be linear.
    In~\citep{Alsing:2018eau}, the likelihood does not need to be known at all, firstly summarising simulations of real data heuristically and then compressing these summaries using an appropriate likelihood in the same way as~\citep{Alsing:2017var}.
	Although a powerful technique, the first step in~\citep{Alsing:2018eau} can potentially be lossy and the likelihood in the second step should be well known to achieve optimal compression of the first step summaries.
    The information maximising neural network can replace both steps in~\citep{Alsing:2018eau} by taking the raw data and providing non-linear, likelihood-free summaries directly from the simulations.
    Likewise, and perhaps more conveniently, the network introduced here is ideally placed to squeeze additional information out of the data after all of the more obvious summaries, such as the power spectrum, have been exhausted.

	\medskip
	In this paper, we have focussed on a few test models used to illustrate the method and its abilities.
    The first set of tests use the network to find a summary of Gaussian signal, without noise, with known noise variance and with unknown noise variance.
    This is a useful example since it can be solved analytically and linear compression, such as MOPED would fail to provide useful summaries of the data.
	We showed that PMC-ABC is able to recover the analytic posterior distribution for the variance of the Gaussian noise nearly exactly, which means that the network has correctly learned the sufficient statistic for this problem.
    It is useful to consider  variance inference as there are many examples in astronomy and cosmology where the variance is informative about the underlying parameters.
    Although the details of the input data and  simulations will be  more complex, variance estimation appears in cases such as estimating the value of the optical depth to reionisation, $\tau$, and recovering B-mode polarisation from probes of the large-angle cosmic microwave background polarisation anisotropies.
    
    \medskip
    Following the success of the first set of tests, the next two examples show further tests on astronomically motivated problems.
    The first shows how extremely noisy raw data can be directly input to the network to constrain cosmological parameters and the second shows how using non-linear summaries are suited to situations where linear summaries can be misleading.
    
	\medskip
   Information maximising neural networks are designed to deal with raw data. We can see IMNNs being useful, or even essential, when trying to calculate posterior distributions of model parameters where the likelihood, describing the distribution of some large number of data points, is unknown. 
    For example, the raw data from large scale structure surveys is infeasibly large. Even the number of summary statistics is $\sim10^4$ and a likelihood cannot be written to describe the physics, the selection bias and the instrument---but the data \textit{can} in principle be simulated from initial conditions.
      The IMNNs presented in this paper to illustrate and explore the concept used a fully connected architecture.  When considering very large data sets we will need to consider network architectures that are adapted to the problem at hand and computationally efficient. 
    For example, assuming the isotropy of the universe transverse to the line of sight, whilst looking radially in redshift space suggests that stacks of convolutional neural networks could be used to deal with raw LSS data.
    As long as patches of the large scale structure (and the instrument) can be simulated to train the convolutional filter, IMNNs should make it possible to extract cosmologically interesting information directly from the raw data---automatically.

\medskip
    The data and original code used in this paper is available at \url{https://doi.org/10.5281/zenodo.1175196}.
    For up-to-date code and current development please use \url{https://github.com/tomcharnock/information_maximiser}.

\medskip
    \section*{Acknowledgements}
    We would like to thank the referee for their excellent comments and useful input.
    This work was supported by the ANR BIG4 grant ANR-16-CE23-0002 of the French Agence Nationale de la Recherche as well as the Simons Foundation and we acknowledge that the work has been done within the Labex ILP (reference ANR-10-LABX-63) part of the Idex SUPER, and received financial state aid managed by the Agence Nationale de la Recherche, as part of the programme Investissements d'avenir under the reference ANR-11-IDEX-0004-02. 
    \bibliography{fisher}
\end{document}